\documentclass[sigplan,screen,10pt,dvipsnames,svgnames,x11names,table,tempa,nonacm]{acmart}
\renewcommand\footnotetextcopyrightpermission[1]{}
\usepackage[T1]{fontenc}
\usepackage{fancyhdr}
\usepackage{mathptmx}
\usepackage[utf8]{inputenc}
\usepackage{times}
\usepackage{framed}
\definecolor{shadecolor}{HTML}{F3F3F3}
\usepackage[inline]{enumitem}
\usepackage[abbreviations]{foreign}  % For \ie, \eg \etc
\usepackage{graphicx} % to include graphics
\usepackage[font={small,stretch=1.0},skip=2pt,belowskip=2pt,labelfont=bf]{caption} % to have figure caption smaller

\usepackage{epsfig}
\usepackage{booktabs} % For formal tables
\usepackage[T1]{fontenc} % fixes issues with urls
\usepackage{times}
\usepackage{pifont}
\usepackage{algorithm}
\usepackage{algpseudocode}
\usepackage{listings}
\usepackage{xspace}
\usepackage{ifthen}
\usepackage{amsmath,bm}
\usepackage{array}
\usepackage{setspace}
\usepackage{tabularx}
\usepackage{units} % changes font style for units
\usepackage{framed}
\usepackage{ragged2e}
\usepackage{flushend}
\usepackage[indentfirst=false,leftmargin=0.5pt,rightmargin=0.5pt,vskip=0pt]{quoting}
 
\usepackage[acronym]{glossaries}
\newacronym{sgx}{SGX}{Intel software guard extensions}
\newacronym{txt}{TXT}{Intel trusted execution technology}
\newacronym{tee}{TEE}{trusted execution environment}
\newacronym{kms}{KMS}{key management system}
\newacronym{epc}{EPC}{enclave page cache}
\newacronym{tls}{TLS}{transport layer security}
\newacronym{os}{OS}{operating system}
\newacronym{drtm}{DRTM}{dynamic root of trust for measurements}
\newacronym{srtm}{SRTM}{static root of trust for measurements}
\newacronym{crtm}{CRTM}{core root of trust for measurements}
\newacronym{rom}{ROM}{read-only memory}
\newacronym{ram}{RAM}{random-access memory}
\newacronym{cpu}{CPU}{central processing unit}
\newacronym{nvram}{NVRAM}{non-volatile random-access memory}
\newacronym{dram}{DRAM}{dynamic random-access memory}
\newacronym{ek}{EK}{endorsement key}
\newacronym{aik}{AIK}{attestation key}
\newacronym{ca}{CA}{certificate authority}
\newacronym{tpm}{TPM}{trusted platform module}
\newacronym{dtpm}{dTPM}{discrete TPM chip}
\newacronym{ftpm}{fTPM}{firmware TPM}
\newacronym{vtpm}{vTPM}{virtual TPM}
\newacronym{itpm}{iTPM}{integrated TPM}
\newacronym{pch}{PCH}{platform controller hub}
\newacronym{tcg}{TCG}{Trusted Computer Group}
\newacronym{pcr}{PCR}{platform configuration register}
\newacronym{pcrd}{dynamic PCR}{dynamic PCR}
\newacronym{pcrs}{static PCR}{static PCR}
\newacronym{ptt}{PTT}{Intel platform trusted technology}
\newacronym{uefi}{UEFI}{unified extensible firmware interface}
\newacronym{bios}{BIOS}{basic input/output system}
\newacronym{pxe}{PXE}{preboot execution environment}
\newacronym{svm}{SVM}{Secure Virtual Machine}
\newacronym{ima}{IMA}{integrity measurement architecture}
\newacronym{vpn}{VPN}{virtual private network}
\newacronym{daa}{DAA}{direct anonymous attestation}
\newacronym{loc}{LOC}{lines of code}
\newacronym{sloc}{SLOC}{source lines of code}
\newacronym{mee}{MEE}{memory encryption engine}
\newacronym{ias}{IAS}{Intel attestation service}
\newacronym{mrenclave}{MRENCLAVE}{enclave hash measurement}
\newacronym{acs}{IBM ACS}{IBM TPM attestation client-server}
\newacronym{cit}{Intel CIT}{Intel open cloud integrity technology}
\newacronym{initramfs}{initramfs}{initramfs}
\newacronym{vm}{VM}{virtual machine}
\newacronym{iaas}{IaaS}{Infrastructure-as-a-Service}
\newacronym{maas}{MaaS}{Metal-as-a-Service}
\newacronym{lpc}{LPC}{low pin count}
\newacronym{me}{CSME}{Intel converged security and manageability engine}
\newacronym{toctou}{TOCTOU}{time of check to time of use}
\newacronym{itl}{ITL}{Invisible Things Lab}
\newacronym{smm}{SMM}{system management mode}
\newacronym{dma}{DMA}{direct memory access}
\newacronym{tcb}{TCB}{trusted computing base}
\newacronym{bmc}{BMC}{baseboard management controller}
\newacronym{ipmi}{IPMI}{intelligent platform management interface}
\newacronym{nic}{NIC}{network interface card}
\newacronym{ssh}{SSH}{secure shell}
\newacronym{hsm}{HSM}{hardware security module}
\newacronym{vmm}{VMM}{virtual machine monitor}
\newacronym{kvm}{KVM}{kernel-based virtual machine}
\newacronym{qemu}{QEMU}{quick emulator}
\newacronym{mktme}{MKTME}{multi-key total memory encryption}
\newacronym{tme}{TME}{Total Memory Encryption}
\newacronym{isp}{ISP}{internet service provider}
\newacronym{ssl}{SSL}{secure sockets layer}
\newacronym{ecc}{ECC}{elliptic-curve cryptography}
\newacronym{rsa}{RSA}{Rivest Shamir Adleman}
\newacronym{vmbr}{VMBR}{virtual-machine based rootkit}
\newacronym{aes}{AES}{advanced encryption standard}
\newacronym{sev}{AMD SEV}{AMD secure encrypted virtualization}
\newacronym{ip}{IP}{internet protocol}
\newacronym{dns}{DNS}{domain name system}
\newacronym{mitm}{MitM}{man-in-the-middle}
\newacronym{arp}{ARP}{address resolution protocol}
\newacronym{tcp}{TCP}{transmission control protocol}
\newacronym{mc}{MC}{monotonic counter}
\newacronym{mcs}{MCS}{monotonic counter service}
\newacronym{rest}{REST}{representational state transfer}
\newacronym{api}{API}{application programming interface}
\newacronym{cve}{CVE}{common vulnerabilities and exposures}
\newacronym{sriov}{SR-IOV}{single root input/output virtualization}
\newacronym{vtd}{VT-d}{Intel virtualization technology for directed I/O}
\newacronym{ecdsa}{ECDSA}{elliptic curve digital signature algorithm}
\newacronym{pal}{PAL}{piece of application logic}
\newacronym{iommu}{IOMMU}{input–output memory management unit}
\newacronym{cicd}{CICD}{continuous integration and continuous deployment}
\newacronym{http}{HTTP}{hypertext transfer protocol}
\newacronym{pax}{PAX}{portable archive exchange}
\newacronym{cdn}{CDN}{content delivery network}
\newacronym{poc}{PoC}{proof-of-concept}
\newacronym{tc}{TC}{trusted computing}
\glsdisablehyper
\makeglossaries

\newcommand\YAMLcolonstyle{\color{red!70!black}}
\newcommand\YAMLkeystyle{\color{black}}
\newcommand\YAMLvaluestyle{\color{blue!70!black}}
\newcommand\YAMLcommentstyle{\color{blue!50!black}}

\makeatletter

\newcommand\language@yaml{yaml}
\expandafter\expandafter\expandafter\lstdefinelanguage
\expandafter{\language@yaml}
{
	keywords={true,false,null,y,n},
	keywordstyle=\color{darkgray}\bfseries,
	basicstyle=\linespread{0.75}\small\ttfamily\color{black},
	sensitive=false,
	comment=[l]{\#},
	morecomment=[s]{/*}{*/},
	commentstyle=\YAMLcommentstyle,
	stringstyle=\YAMLvaluestyle,
	moredelim=[l][\color{orange}]{\&},
	moredelim=[l][\color{magenta}]{*},
	moredelim=**[il][\YAMLcolonstyle{:}\YAMLvaluestyle]{},
	morestring=[b]',
	morestring=[b]",
	literate = {---}{{\ProcessThreeDashes}}3
	{>}{{\textcolor{red!70!black}\textgreater}}1
	{|}{{\textcolor{red!70!black}\textbar}}1
	{-}{{\mdseries-}}1,
}

\lst@AddToHook{EveryLine}{\ifx\lst@language\language@yaml\YAMLkeystyle\fi}
\makeatother
\usepackage{array}
\usepackage{txfonts}
\newcommand{\ftextnumero}{{\fontfamily{txr}\selectfont \textnumero}}

\usepackage{url}

\usepackage{breakurl}
\hypersetup{
  colorlinks,
  linkcolor={green!50!black},
  citecolor={red!70!black},
  urlcolor={blue!70!black}
}

\newcommand{\myparagraph}[1]{\vspace{1mm} \smallskip \noindent{\bf {#1}}}

\usepackage{newfloat}
\DeclareFloatingEnvironment[fileext=lol,name=Listing]{listing}
\usepackage{subcaption}
\DeclareCaptionSubType{listing}
\usepackage{todonotes}
 
\newboolean{showcomments}
\setboolean{showcomments}{true}
\ifthenelse{\boolean{showcomments}}
{ \newcommand{\mynote}[3]{
    \fbox{\bfseries\sffamily\footnotesize#1}
    {\small$\blacktriangleright$\textsf{\emph{\color{#3}{#2}}}$\;\blacktriangleleft$}}}
    { \newcommand{\mynote}[3]{}}
\newcommand{\sys}{TSR\xspace}
\newcommand{\sysfull}{trusted software repository\xspace}

\pdfoutput=1

\makeatother
\usepackage{tcolorbox}
\title{A practical approach for updating an integrity-enforced operating system}

\author{Wojciech Ozga}
\affiliation{%
  \institution{TU Dresden, Germany}
}

\author{Do Le Quoc}
\affiliation{%
  \institution{TU Dresden, Germany}
}

\author{Christof Fetzer}
\affiliation{%
  \institution{TU Dresden, Germany}
}

\date{}


\begin{abstract}
Trusted computing defines how to securely measure, store, and verify the integrity of software controlling a computer.
One of the major challenge that make them hard to be applied in practice is the issue with software updates.
Specifically, an operating system update causes the integrity violation because it changes the well-known initial state trusted by remote verifiers, such as integrity monitoring systems. 
Consequently, the integrity monitoring of remote computers becomes unreliable due to the high amount of false positives.

We address this problem by adding an extra level of indirection between the operating system and software repositories.
We propose a \sysfull (\sys), a secure proxy that overcomes the shortcomings of previous approaches by \emph{sanitizing} software packages. 
Sanitization consists of modifying unsafe installation scripts and adding digital signatures in a way software packages can be installed in the operating system without violating its integrity. \sys leverages shielded execution, \ie, Intel SGX, to achieve confidentiality and integrity guarantees of the sanitization process. 

\sys is transparent to package managers, and requires no changes in the software packages building and distributing processes.
Our evaluation shows that running \sys inside SGX is practical; since it induces only $\sim1.18\times$ performance overhead during package \emph{sanitization} compared to the native execution without SGX.
\sys supports $99.76$\% of packages available in the main and community repositories of Alpine Linux while increasing the total repository size by $3.6$\%.

\end{abstract}

\begin{document}
\thispagestyle{plain}
\thispagestyle{fancy}
\maketitle
\errorstopmode
%!TEX root = paper.tex
\section{Introduction}
\label{sec:introduction}
% trusted computing \cite{trusted_computing_2009, tcg_org, Pearson:2002:TCP:1088853}
In the last years, \gls{tc} technologies, such as \gls{txt}~\cite{intel_txt_whitepaper}, \gls{ima} \cite{tcg_ima_spec, ima_design_2004}, and \gls{tpm}~\cite{tpm_2_0_spec_architecture,ibm_tpm_tss}, have received much attention both in industry and academia because of their capacities for measuring integrity, remote attestation, and sealing.
% Typically, the \gls{tc} technologies rely on cryptographic hash mechanisms to provide integrity guarantees for applications deployed on an untrusted computing platform.
%regardless of the trustworthiness of the underlying system software. 
While promising at first glance, the approach of leveraging \gls{tc} technologies suffers from technical issues.
One of the major problems of applying them in production systems is the lack of support for \gls{os} updates. 
Specifically, the security patches, which might be released frequently and installed automatically, break system integrity.
We refer to integrity as a security property describing that a computer runs only expected software in the expected configuration.

% We refer to system integrity as a security property describing computing systems running legitimate software in the expected configuration.
% The system integrity is a security property used by \emph{verifiers} (\eg, monitoring systems \cite{intel_secl, opencit_01_org, ibm_tpm_acs}, \glsdesc{vpn} access points \cite{strongswan_org}) to identify compromised or misconfigured systems. 
% Such systems rely on hardware and software technologies to collect and provide measurement reports containing integrity digests of every configuration file, dynamic library, script, and executable that has been loaded to the memory since the computer boot. 
% The verifiers ensure that the measurement report consists of only legal measurements that indicate that the remote platform is controlled by legitimate software with the expected configuration. 
% However, it is challenging and barely practical to maintain such a list \cite{scalable_attestation}. 
% Before installing or updating any software, the list must be extended with measurement digests of all files that would change with the update. 
% \cite{turtles_mccune}

\begin{figure}[tbp!]
\centering
% \captionsetup{skip=4pt}
\includegraphics[width=0.48\textwidth]{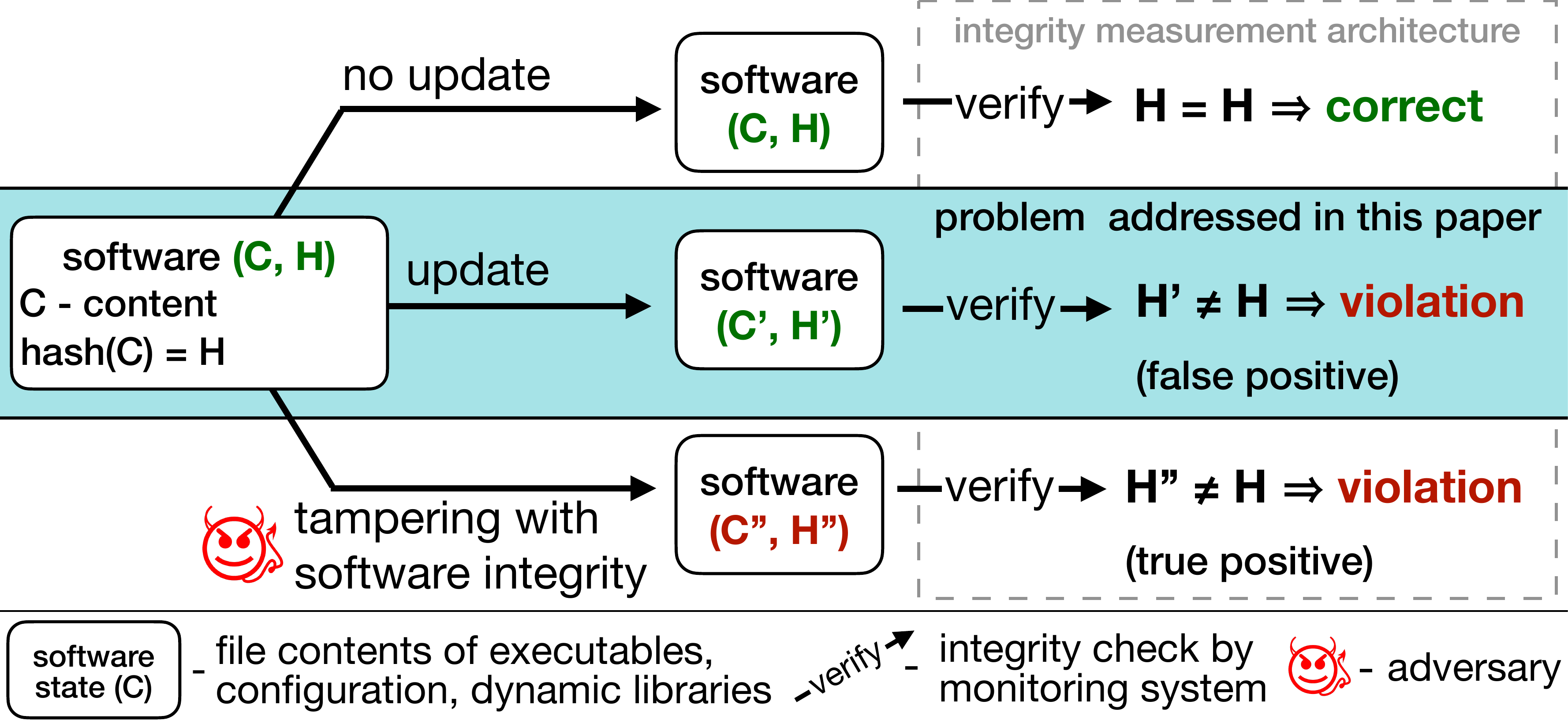}
\caption{
    Problem of installing software updates in an integrity-enforced OS.
    Software updates change software integrity measurement, which is reported by the monitoring systems as integrity violation.
    The main question addressed in this paper: How to distinguish between software manipulated by an adversary and correctly updated software?
}
\label{fig:problem_overview}
\end{figure}

To illustrate the problem of installing software updates, we first describe the concept of integrity verification provided by \gls{tc} technologies. Verifiers (\eg, monitoring systems \cite{intel_secl, opencit_01_org, ibm_tpm_acs} or \glsdesc{vpn} access points \cite{strongswan_org}) use hardware and software technologies \cite{intel_txt_whitepaper, ima_design_2004, intel_ptt_whitepaper_2014}, which implement trusted computing \cite{drtm_tcg, tpm_2_0_spec, tcg_ima_spec}, to identify compromised (executing not allowed software) or misconfigured (having not permitted configuration) systems.
In more detail, verifiers read from a remote computer a list of cryptographic hashes (measurement report) calculated over every file loaded to the computer memory since the computer boot.
Verifiers detect integrity violations by comparing hashes to a whitelist, which is a list that contains hashes of approved software and configuration.
% It is possible because a cryptographic hashes are unique for every file content.
Unfortunately, verifiers cannot distinguish whether software integrity changed due to malicious behavior or a legitimate software update (see \autoref{fig:problem_overview}).

Berger et al. 2015 \cite{scalable_attestation} proposed to include in the measurement report digital signatures, which certify the integrity hashes of trusted software.
The approach simplifies the verification process because verifiers require only a single certificate to check the signatures instead of a whitelist of all possible cryptographic hashes. 
Consequently, it opened an opportunity to support OS updates because the updates could incorporate digital signatures to vouch for the integrity of files changed during the update.
OS distributions would have to change their software packaging process to issue and to insert digital signatures of files inside packages \cite{imasig_updates}. 
This approach has, however, two limitations, which we address in this paper. 
First, it requires changes to the existing procedures of creating packages for every OS distribution.
Second, software packages contain not only files that are extracted to the filesystem but also configuration scripts that might alter OS configuration, thus breaking the integrity.

Instead of modifying the well-established process of package generation (which requires approval from the entire open-source community), an alternative approach consists of creating a standalone repository with modified packages containing digital signatures \cite{imasig_updates}.
The approach requires a trusted organization which owns a signing key and re-creates packages after injecting digital signatures.
Such an organization must put additional efforts to protect the signing key and must have a good reputation to convince users to trust it.
We argue that it might be difficult to achieve, considering incidents from the past, when signing keys of major Linux distribution were leaked affecting millions of users \cite{fedore_signingkey_compromised, redhat_ssh_signing}.

Another problem is that an adversary controlling a repository can provide the OS with outdated packages containing known vulnerabilities (replay attack), or even prevent the OS from seeing the update (freeze attack) \cite{cappos_look_2008, cappos2008package}.
The secure choice is to rely only on the \emph{original repository}, which is a repository managed by a trusted organization, such as an official software repository of the OS distribution.
But, this approach does not tolerate the original repository failure, thus the OS must also accept \emph{mirrors}.
Mirrors store a copy of the original repository, and, in the case of open-source distributions, are hosted voluntarily.
As reported by previous studies~\cite{cappos_look_2008}, it is not difficult to create a custom mirror that becomes accepted as an official mirror.
Therefore, we must tolerate that some of the available mirrors are controlled by an adversary, exposing operating systems to threats mentioned above.
For example, it happened that a compromised mirror of a popular repository distributed a vulnerable version of the software, allowing an adversary to remotely access the system \cite{compromised_mirror}.

We present the \sysfull (\sys), an intermediate layer between the OS and the software repository that provides \emph{sanitized} software packages. 
The installation of sanitized packages causes deterministic changes to the OS configuration and filesystem.
Because such changes are verifiable by monitoring systems, \sys eliminates the risk of false-positives.
According to our measures, sanitization enables 99.76\% of packages available in the Alpine main and community repositories to be safely installed in integrity-enforced operating systems.

\sys requires zero code changes to both monitoring systems as well as operating systems.
Due to the shared nature of the software repositories, we designed \sys as a service that can be hosted on the third-party resources, \ie, in the cloud. 
\sys exploits \gls{tee}, \ie, \gls{sgx} \cite{mckeen_innovative_2013, costan2016intel, anati2013innovative}, to protect the signing keys and \sys integrity.
Our evaluation shows that running \sys inside \gls{sgx} is practical; SGX induces in average $1.18\times$ performance overhead during sanitization, up to $1.96\times$ for packages exceeding available SGX memory.
Note that the sanitization is performed in batch mode and hence, the slowdown has no practical impact.

Last but not least, \sys accepts security policies, which reflect organizational-specific security requirements.
Specifically, each organization defines a list of mirrors.
\sys uses mirrors to establish quorum on the correct version of a software package, thus tolerating mirrors compromised by an adversary.
We show that \sys requires up to 2.2 seconds to establish a quorum from official Alpine mirrors distributed over three continents.

\begin{figure}[tbp!]
\centering
% \captionsetup{skip=4pt}
\includegraphics[width=0.48\textwidth]{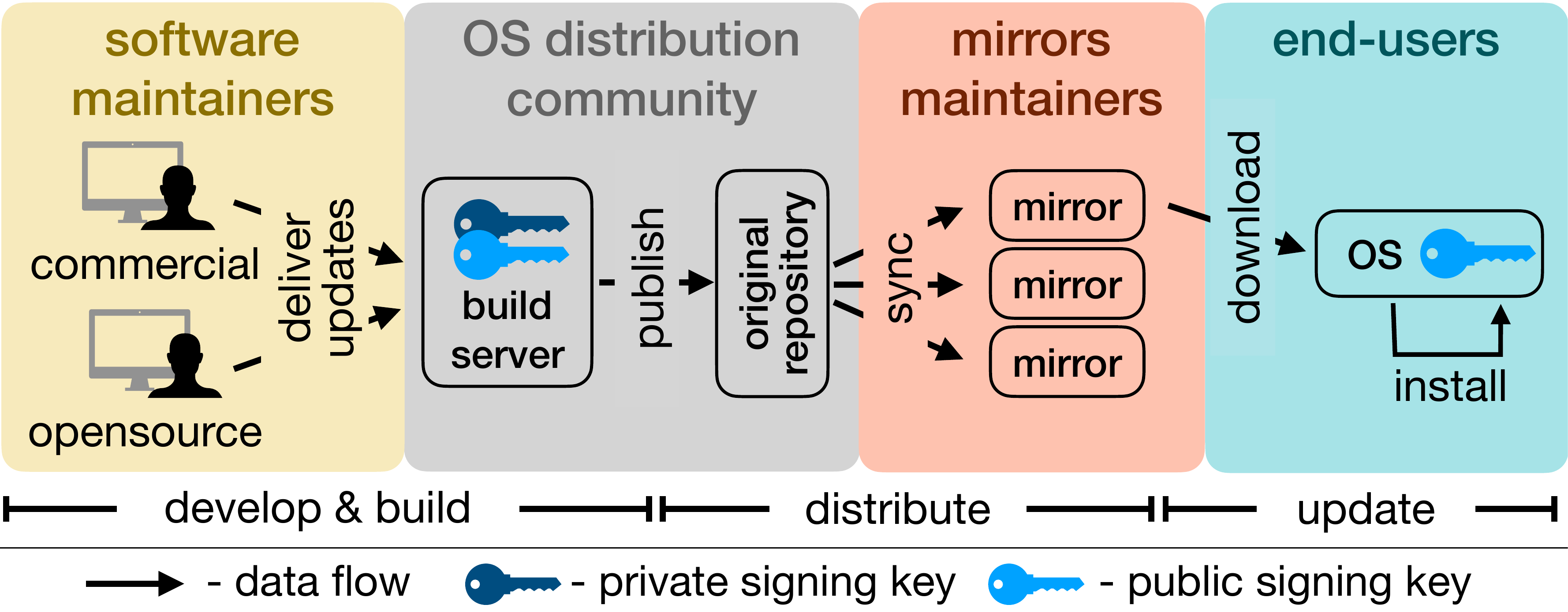}
\caption{
    Overview of software update process. Colors indicate different administrative domains and are consistent across all figures.
}
\label{fig:update_process_overview}
\end{figure}

In summary,  we make the following main contributions:
\begin{enumerate}[nosep,leftmargin=*]
\item We propose a practical solution to support OS updates in integrity-enforced systems, with the following properties:
	\begin{enumerate}[nosep,leftmargin=*, label=\emph{(\alph*)}]
	\item The software packages are safe to install in integrity-enforced operating systems (\S\ref{sec:design_sw_updates}).
	\item Our solution is transparent to the existing software update processes and infrastructure (\S\ref{sec:design_community}).
	\item A minority of mirrors exhibiting Byzantine behavior are tolerated (\S\ref{sec:byzantine_mirrors}).
\end{enumerate}
\item We realize the above-mentioned design by developing \sys  --- a secure proxy framework for supporting software updates in integrity-enforced operating systems (\S\ref{sec:implementation}).
\item We have evaluated \sys using a series of micro-benchmarks, and a real-world use case --- Alpine Linux package updates (\S\ref{sec:evaluation}).
%\item \sys, implementation of the software repository implementing the above-mentioned design supporting Alpine Linux (\S\ref{sec:implementation}).
%\item Evaluation of the system (\S\ref{sec:evaluation}).
\end{enumerate}

%!TEX root = paper.tex
\section{Background}
\label{sec:background}
To better understand the decisions taken in designing \sys, we start by providing background information on software update processes and about existing technologies used to collect, report, and verify system integrity. 
% First, we give a high-level overview of the software update process.
% Then, we explain the idea of \texttt{chain of trust} \cite{wilkins_secure_boot_2013} that, together with the \texttt{hardware root of trust} permits building tamper proof integrity measurements list.

\subsection{OS updates}
\autoref{fig:update_process_overview} shows a high-level overview of an OS update process: releasing, exposing, and installing new software versions.
The process begins when software maintainers create a new software release that contains bug fixes or new features.
The \gls{os} distribution community uses the source code of the new software release to create a software package.
A software package is an archive containing software-specific files and meta-information required by the \gls{os} to install and manage the package.
Packages are stored in a repository, from which end-users download them.
A repository stores also a \emph{metadata index} that contains a digitally signed list of all packages.
In this paper, we refer to a software repository controlled by an OS distribution community as an \emph{original repository}.
The original repository is a root of trust for software updates.
The metadata file downloaded from the original repository provides information about the most recent versions of software available in the repository. 
As such, it can be used to verify that the OS is up-to-date.

Repository mirrors contain a copy of the original repository. 
They are used to distribute the load and to decrease the latency of downloading packages.
The community has limited control over the mirrors, which are typically supported by volunteer organizations.
Importantly, mirrors do not have access to the signing key.
End-users verify that the metadata file and packages downloaded from mirrors originate from the original repository by verifying digital signatures using a public portion of the signing key provided by the OS distribution community.

% A subset of updates, called security updates, are released to fix bugs and vulnerabilities.
% It is a widely accepted practice to let operating systems automatically install security updates.
% \wojciech{provide info about existing tools to proof that it works}.

% Package manager is a software running on the OS.
% It communicates with software repository to read a \emph{repository index file}.
% A repository index file enumerates software (including its versions, dependencies) available in the repository.
% The package manager uses it to learn which packages can be updated by comparing the currently installed versions with versions available in the index.
% Eventually, the package manager downloads and installs newer versions of packages.

\subsection{Package managers}
Operating systems use \emph{package managers} to simplify installation, update, and removal of software. 
The majority of distributions ship with package managers that use pre-built packages (\eg, .rpm, .deb \cite{DebianPackageSystem}, .apk \cite{AlpinePackageManagement}), but some build software directly from sources \cite{GentooPortage, ArchBuildSystem}. 
In this paper, we focus only on the pre-built packages, which we refer further as \emph{packages}.

\begin{figure}[tbp!]
\centering
% \captionsetup{skip=4pt}
\includegraphics[width=0.48\textwidth]{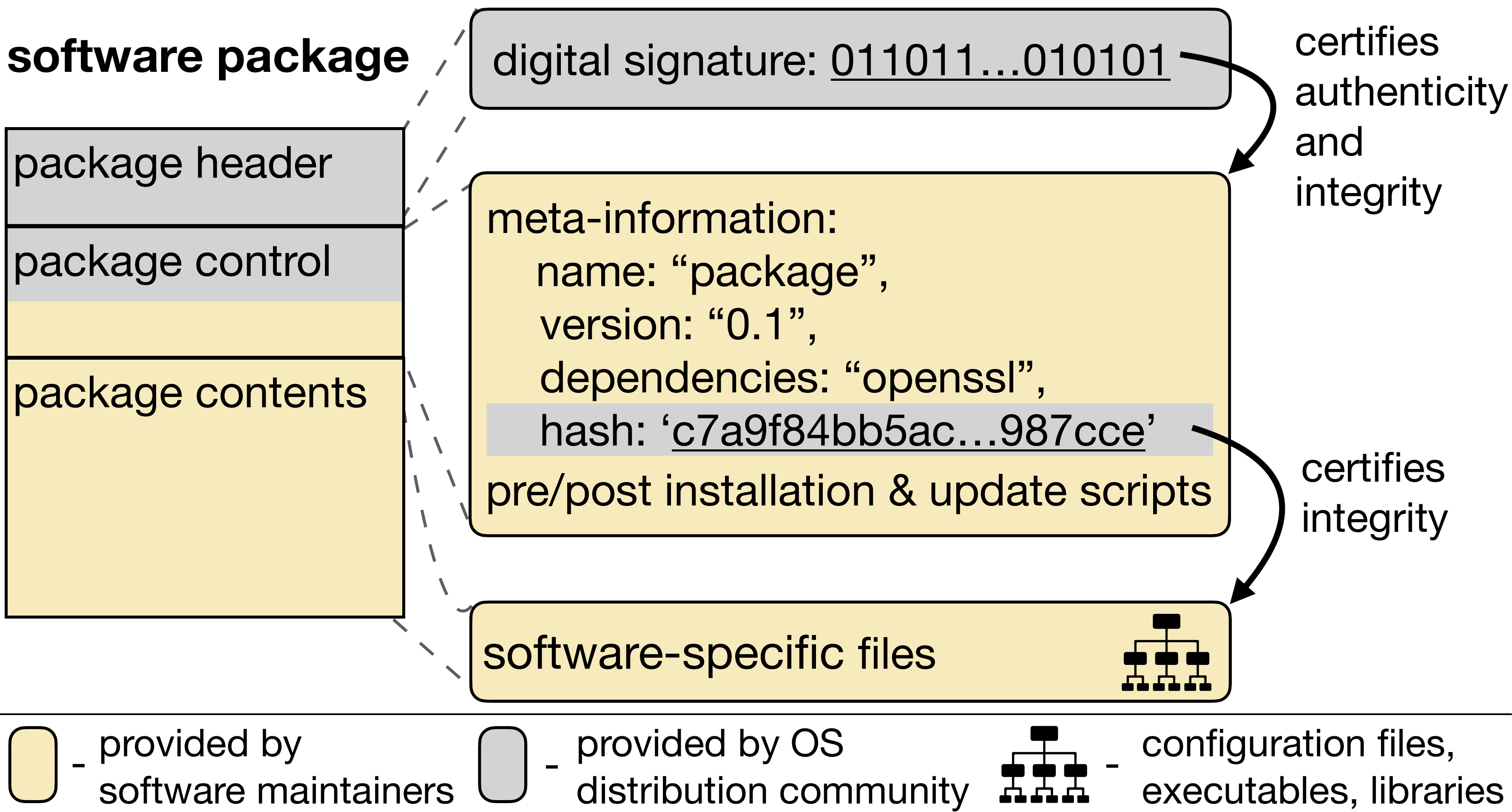}
\caption{
    The internal structure of a software package, \ie, Alpine APK package format. 
    The package authenticity and integrity can be verified by using the digital signature and the content hash.
    The digital signature is stored inside the header, and is issued over the package control. 
    The hash of the package contents is stored inside the meta-attributes of the package control.
}
\label{fig:package_structure}
\end{figure}

A package is an archive containing software-specific files, installation scripts, meta-information (such as dependency on other packages), and digital signatures.
\autoref{fig:package_structure} shows an example of a package in the Alpine Linux .apk format.
The package header stores a digital signature issued by a developer with an offline signing key (a private key stored off the repository).
The digital signature permits verifying the authenticity and the integrity of the package control, which contains installation scripts and meta-information describing package dependencies, software version, and a cryptographic hash of package contents.
The hash permits verifying the integrity of executables, dynamic libraries, and configuration files stored inside the package.

To install the package, the package manager first downloads it from the repository, or from middlemen such as a \gls{cdn} or mirrors.
After that, it verifies that a trusted entity created the package. 
Finally, it runs installation scripts and extracts software-specific files to the file system.

%  shows an Alpine APK package format. The package is a file created by the concatenation of three archives. The first archive contains a digital signature over the second archive. The second archive contains configuration scripts and the metafile with package-specific information, such as package name, version, dependencies, the integrity digest of the third archive. The third archive contains files that are copied to the destination file system during the installation.

% Obtaining an up-to-date repository index 
% To distribute load and decrease latency, distributions use several servers (called mirrors) containing a copy of the main repository. 

\subsection{Integrity measurements}
% The first component, typically a static, immutable piece of code \cite{choinyambuu_crtm_2011}), initiates chain of trust.
To bootstrap the computer, multiple low-level software components execute.
They form a chain of trust by following the rule that every component calculates a cryptographic hash (an integrity measurement) of the next component before executing it. 
The measurements are stored in tamper-resistant memory of a hardware root of trust, \eg, \gls{tpm} \cite{tpm_2_0_spec}.
Eventually, one of the components measures the bootloader, which measures and loads the kernel.
At the kernel level, the integrity measurements continue.
The Linux kernel integrity measurement subsystem (Linux \gls{ima} \cite{tcg_ima_spec, ima_design_2004}) measures each file, executable, or library before loading it to the memory. 
The list of all measurements, certified by a hardware root of trust (\eg, \gls{tpm} \cite{tpm_2_0_spec}), vouches for the system integrity \cite{tcg_tpm_attestation}.
Integrity monitoring systems use the measurements to verify if only expected software executed on the computer since its bootstrap.

% Trusted computing specifies an alternative leveraging hardware extensions of modern \glspl{cpu}  in  or by \glsdesc{drtm} \cite{drtm_tcg} that leverages hardware extensions of modern \glspl{cpu} \cite{intel_txt_whitepaper, amd_svm_2005}. 
% In the case of Linux systems, the Linux IMA \cite{ima_design_2004}, which is the \gls{ima} \cite{tcg_ima_spec, ima_design_2004} implementation in the Linux kernel \cite{linuxkernel}, performs measurement and appraisal of files loaded to the memory. 
% All measurements are sent to the \gls{tpm} \cite{tpm_2_0_spec}, which provides tamper-resistant storage. 
% The TPM offers a remote attestation protocol \cite{tcg_tpm_attestation} in which the TPM generates a digitally signed quote certifying \glspl{pcr}. 
% The quote is used by monitoring systems to verify the integrity of the remote platform.

\section{Threats and challenges}
\label{sec:threatmodel}

\subsection{Threat model}
% Software update infrastructure, \ie, original repository, mirrors, is security-critical components providing operating systems with bug fixes and security updates.
We assume an adversary whose goal is to install vulnerable software on a remote computer by exploiting the software update mechanism.
A remote computer is configured to install updates from \sys, which itself relies on the original repository and official mirrors.
An adversary has root access to the machine running \sys and to the minority of machines hosting mirrors.
In more detail, she controls up to \emph{f} mirrors out of a total of \emph{2f + 1} mirrors available to \sys.
The adversary has access to all outdated packages that contain vulnerabilities, including outdated signed metadata files.
By having root access to machines hosting \sys and mirrors, she can prevent network connection to the original repository and arbitrary mirrors.

We assume that the OS distribution community, software maintainers, their internal processes (\ie, software development, packages build), and infrastructure are trusted.
In particular, packages are build using legitimate compilers; signing keys are well protected; the original repository provides the most recent software versions.
We do not consider attacks resulting from the incorrect design of package formats and metadata, \ie, the endless data attack and the extraneous dependencies attack \cite{cappos_look_2008}.
The assumption is practical because main repositories hosted by the popular Linux distributions (\ie, Debian, Ubuntu, RedHat, Alpine) and their corresponding package managers mitigate the attacks by digitally signing the metadata, which also includes packages file sizes and integrity hashes.

% An adversary who takes control of the software repository might cause an OS to install vulnerable versions of packages.
% Especially, she might launch the rollback attack \cite{cappos_look_2008} to expose an outdated, vulnerable version of the software or launch the freeze attack \cite{cappos_look_2008} to prevent the OS from seeing a software update.

% She does not know about any zero-day vulnerabilities.
% (\eg, we exclude \cite{git_fork_attack_2016}) 

% \sys guarantees to serve the most up-to-date packages, assuming different administrative domains control the community repositories and mirrors. 
% It is based on the assumption that the adversary 
% The assumption is practical because although an adversary might leverage her administrative privileges to tamper with network traffic to prevent communication among \sys and certain hosts, she cannot restrict network access between legitimate mirrors and a software repository hosted by a trusted entity.

The \glspl{tee} are vulnerable to side-channel attacks \cite{Kocher2018spectre, vanbulck2018foreshadow}. 
We exclude them from the threat model, assuming they can be addressed using dedicated tools \cite{SpecLH2019, varys_2018, specfuzz_oleksi}, by updating microcode \cite{intel2018l1tf}, or by excluding a particular type of hardware during the remote attestation protocol \cite{johnson2016intel}.
% We also exclude physical and hardware attacks, such as .
% In this paper we focus on building a software repository that provide packages that can be safely installed on the integrity-enforced OS.
% Moreover, the repository guarantees to serve the most up-to-date packages without having control over the community repositories and mirrors.

\subsection{Problem statement}
\begin{figure}[tbp!]
\centering
% \captionsetup{skip=4pt}
\includegraphics[width=0.48\textwidth]{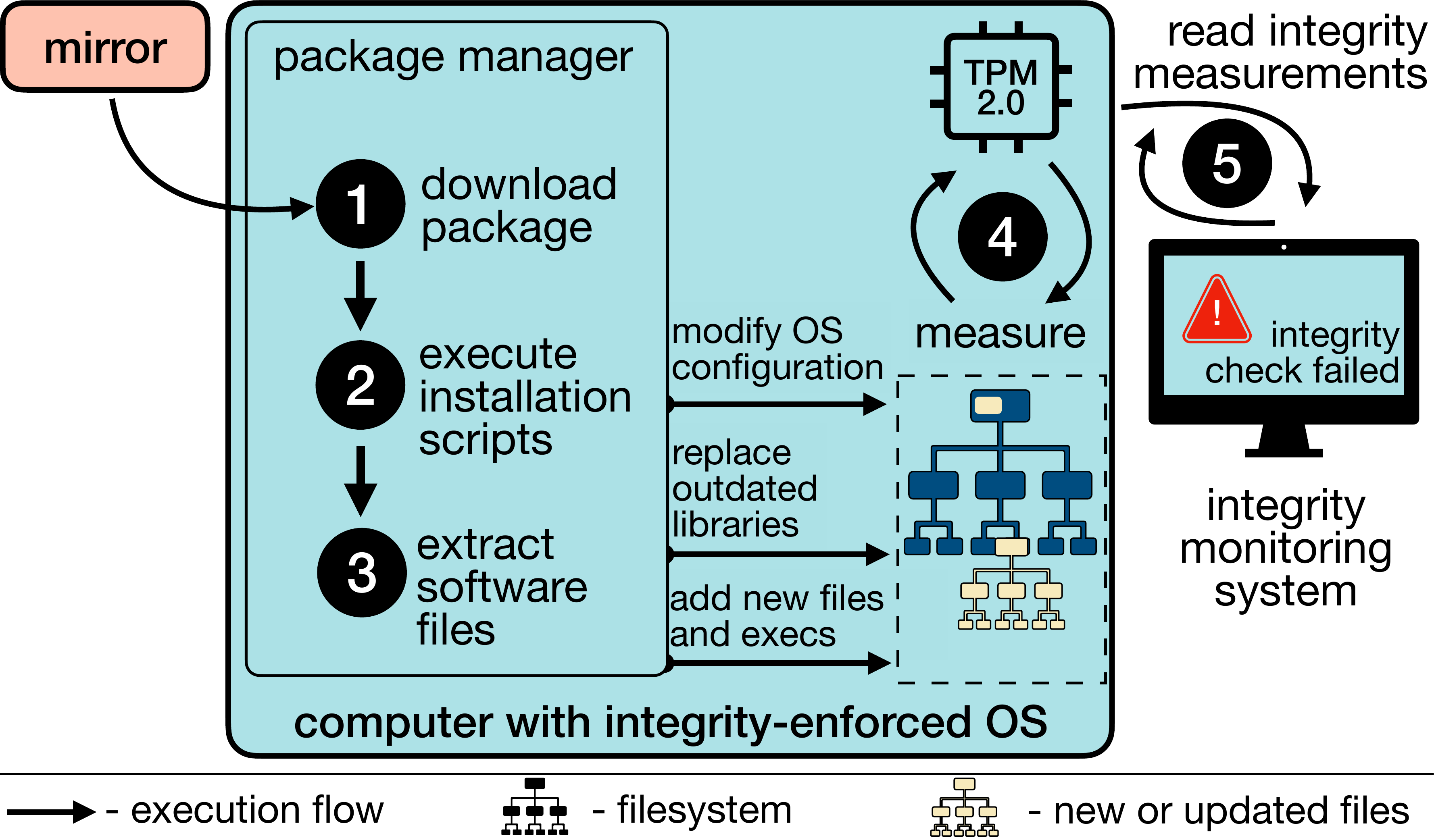}
\caption{
    Example of the package installation that changes the OS configuration and filesystem.
    Monitoring systems consider such a system compromised because the new OS configuration might, for example, allow an adversary to get remote access to the computer or remotely exploit vulnerabilities in the replaced dynamic libraries.
}
\label{fig:update}
\end{figure}

We now introduce the main challenges and problems that shaped the \sys design.

% \myparagraph{Fragile integrity measurements}
%Trusted computing provides a set of hardware and software technologies that allow collecting and reporting integrity measurements, \ie, what configuration files, executables, and dynamic libraries have been loaded to the computer memory since the computer boot. 
% The verifiers, like monitoring systems, consider platform to be compromised if it has loaded to the memory unknown or not properly configured software.
% Even a minor change to the filesystem might cause the machine to be considered compromised.

\begin{shaded*}
\begin{quoting}
\myparagraph{\textbf{Problem 1}}: How to modify the package so that the changes made to the OS configuration and filesystem are verifiable by the monitoring system?
\end{quoting}
\end{shaded*}
The monitoring systems regularly verify that remote computers run only expected software in the expected configuration. 
Machines that fail the attestation might be restarted or reinstalled to bring the system back into the correct state. 
Also, there exist mechanisms to enforce OS integrity locally. 
Such mechanisms are built into the kernel (\eg, IMA-appraisal \cite{ima_appraisal}), allowing the kernel to authorize each file before loading it to the memory.
They make the integrity attestation more robust, preventing accidental or malicious changes to the filesystem.

The main problem of applying trusted computing in production systems is, however, that software updates cannot be safely installed because they modify the OS configuration and change files in a way unknown to monitoring systems. 
\autoref{fig:update} shows why the package installation might move the OS into an untrusted state.
After the package is downloaded (\raisebox{-1pt}{\ding{202}}), the package manager executes software-specific installation scripts that modify the OS configuration (\raisebox{-1pt}{\ding{203}}). 
Moreover, the package manager extracts software-specific files (\raisebox{-1pt}{\ding{204}}), which contents are not known to verifiers.
The integrity of the OS configuration files and software-specific files is measured by trusted computing components (\raisebox{-1pt}{\ding{205}}).
Eventually, a monitoring system uses remote attestation to read the measurements (\raisebox{-1pt}{\ding{206}}), thus detecting the OS integrity change.
The OS is considered compromised.

A strawman approach consists of providing the monitoring system with a list of valid measurements before installing a new package. 
In practice, constructing such a list a priori is a difficult problem because of the complex nature of software dependencies, the OS configuration depending on the order in which software has been installed, and unpredictable schedules of security updates.
% Therefore, we focus on the alternative approach where each package contains digital signatures conforming the integrity of each file \cite{imasig_updates}.
% The problem we address is hence:
% \begin{shaded*}
	
% \end{shaded*}

\begin{figure}[tbp!]
\centering
% \captionsetup{skip=4pt}
\includegraphics[width=0.48\textwidth]{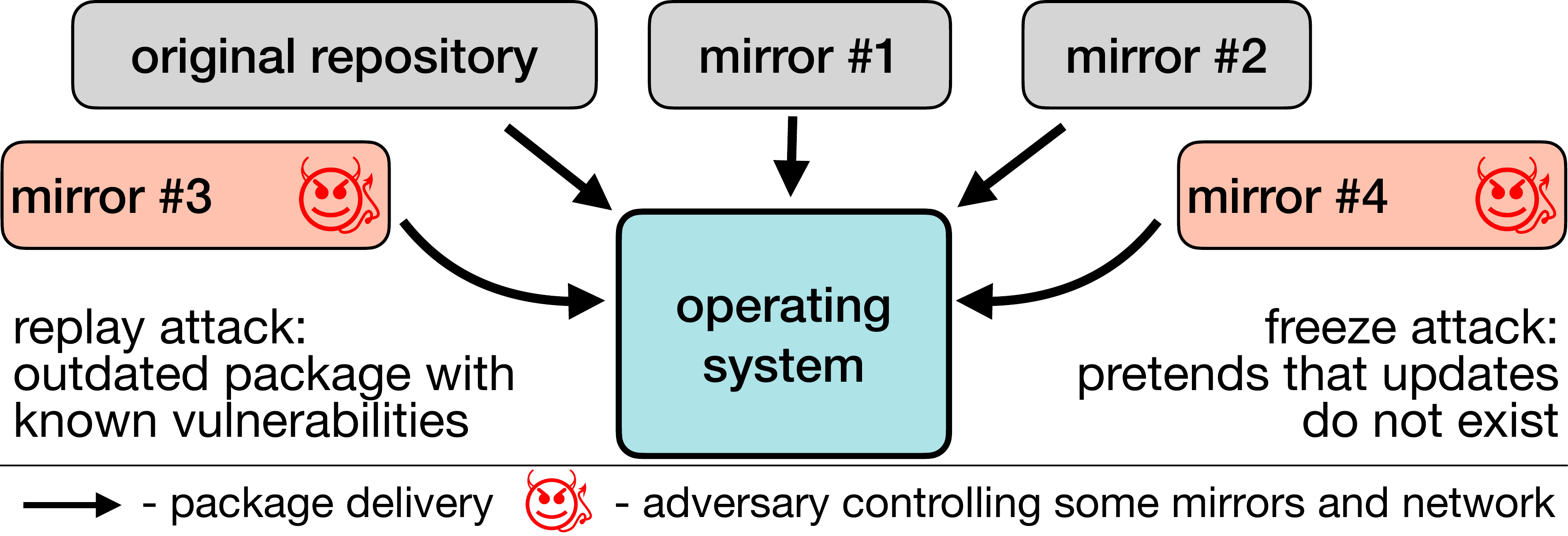}
\caption{
    Mirrors controlled by an adversary can provide outdated packages with known vulnerabilities (replay attack) or completely hide the presence of software updates (freeze attack).
    An adversary might prevent access to the original repository (the root of trust) forcing OS to rely on mirrors.
}
\label{fig:mirrors_malicious}
\end{figure}

\begin{figure*}[btp]
\centering
% \captionsetup{skip=4pt}
\includegraphics[width=\textwidth]{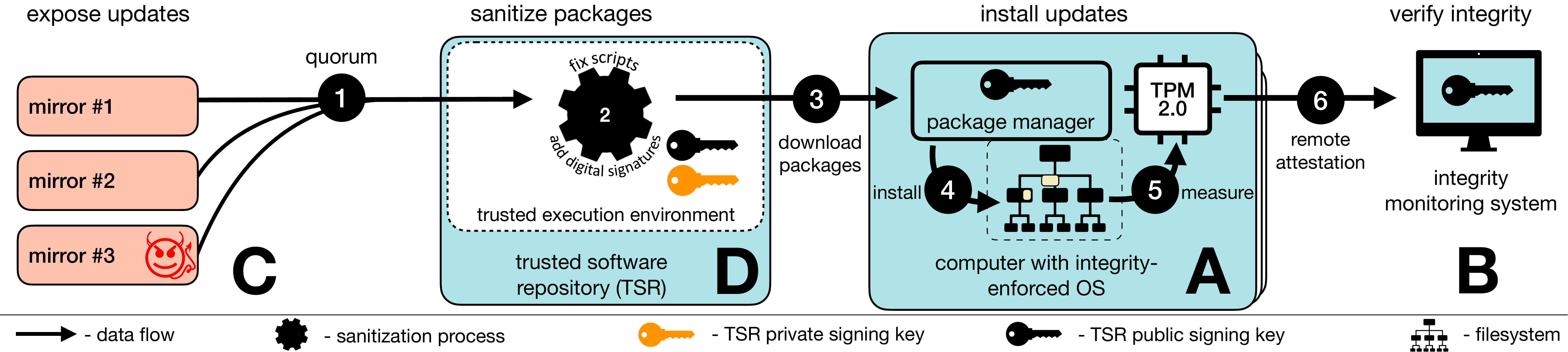}
\caption{
    High-level overview of \sysfull (\sys). \sys is a proxy that modifies packages in a way they are safe to be installed in the integrity-enforced operating systems.
    \sys, TPM, and the integrity monitoring system are trusted.
}
\label{fig:highlevel}
\end{figure*}

\begin{shaded*}
\begin{quoting}
\myparagraph{\textbf{Problem 2}}: How to modify packages without changing the well-established package creation requiring community approval?
\end{quoting}
\end{shaded*}
Previous studies proposed changing the package creation process operated by different Linux communities to include digital signatures that vouch for individual file integrity \cite{scalable_attestation}.
Although different approaches have been proposed \cite{berger_dpkg_patch, garrett_dpkg_patch}, they have not gained enough community approval and have not been merged into upstream repositories.
Therefore, a practical solution should not require changes to the existing package creation processes, thus be transparent to the existing update infrastructure and processes.
% Solving the \textit{problem 1} does not make our solution practical, however.
% We must hence address the problem of: 

% \begin{shaded*}
	
% \end{shaded*}

\begin{shaded*}
\begin{quoting}
\myparagraph{\textbf{Problem 3}}: How to protect the signing key and to guarantee the correct generation of signatures in the presence of a powerful adversary with administrative access to \sys?
\end{quoting}
\end{shaded*}
If we assume that we know how to modify the package (problem 1), the OS would reject the modified package because its digital signature would not match the package contents.
This is expected behavior because it prevents operating systems from installing packages tampered by an adversary.
Therefore, a new package content must be certified again.
However, without community support, it is impossible to issue the signature because the community would restrict access to the signing key (problem 2).

An alternative approach is to let \sys generate a custom signing key, so it uses it to sign all modified packages.
However, an adversary with access to the machine on which the signing keys are used might extract the signing key by simply reading the process memory using administrative rights or by exploiting memory corruption techniques \cite{memory_corruption_techniques}.
Consequently, the adversary might sign arbitrary packages compromising all operating systems that trust the signing key.
% Therefore, we address the following problem:

% \begin{shaded*}
	
% \end{shaded*}

\begin{shaded*}
\begin{quoting}
\myparagraph{\textbf{Problem 4}}: How to ensure access to the most up-to-date packages despite having no connection to the main software repository?
\end{quoting}
\end{shaded*}
Software repositories are maintained by the OS distributions and provide public access to packages and updates.
We refer to such repositories as \emph{original repositories} because new versions of packages and software updates are published directly there.
Although the secure choice would be to always rely on the original repository controlled by a trusted organization, such a decision would introduce a single point of failure.
For this reason, original repositories propagate software updates to mirrors, which expose them to the wide range of end client machines.

%  (\autoref{fig:update_process_overview}).
% Although it is configurable which mirror is used by an OS, a common choice is to choose the fastest one in order to reduce the latency of downloading updates.

As reported by previous studies, an adversary controlling the mirror can serve outdated, vulnerable packages, decreasing the security of operating systems relying on that mirror \cite{cappos_look_2008, cappos2008package}.
\autoref{fig:mirrors_malicious} shows that an adversary might prevent OS from accessing the original repository, and forcing the OS to use mirrors under her control.
% The problem we address is: 

% \begin{shaded*} 
	
% \end{shaded*}

%!TEX root = paper.tex
\section{Approach: Trusted Software Repository}
\label{sec:design}

Our objective is to provide an architecture that:
\begin{enumerate}[label=\emph{\textbullet}]
  \item provides software updates which can be safely installed in an integrity-enforced OS,
  \item requires no changes to the process of how communities create and distribute software packages,
  \item tolerates threats defined in \S\ref{sec:threatmodel}.
\end{enumerate}

\subsection{Design}
\autoref{fig:highlevel} shows a high-level overview of the \sys design.
It consists of four components: 
% Therefore, we propose an alternative approach that does not require changes to the existing well-established package creation process. 
% Our service (\sys), transparent for both operating systems and package creators, sanitizes packages.
\begin{enumerate*}[label=\emph{(\Alph*)}]
  \item an integrity-enforced OS measured by trusted computing components,
  \item a monitoring system which remotely verifies OS integrity,
  \item mirrors, copies of the original repository, containing OS-dependent software packages,
  \item \sys, an intermediate layer that provides the OS with access to software packages that are safe to install in an integrity-enforced OS.
\end{enumerate*}

Now, we present how \sys integrates with the software update process.
First, \sys fetches the most up-to-date packages from mirrors (\raisebox{-1pt}{\ding{202}}) and modifies them in a way they are safe to install (\raisebox{-1pt}{\ding{203}}).
Next, the package manager queries \sys to collect information about the latest versions of packages.
After selecting packages to update, it downloads them from \sys (\raisebox{-1pt}{\ding{204}}).
Then, the package manager installs them (\raisebox{-1pt}{\ding{205}}), causing partial update of the existing OS configuration, replacement of existing files (\eg, dynamic libraries), and extraction of new files into the filesystem.
Trusted computing components regularly measure these changes, and the corresponding integrity measurements are stored inside a \gls{tpm} chip (\raisebox{-1pt}{\ding{206}}).
The monitoring system collects the attestation report (\raisebox{-1pt}{\ding{207}}), which next to integrity measurements, contains the corresponding digital signatures.
After verifying the digital signatures and the integrity measurements, the monitoring system accepts a new state of the updated OS.

\newcommand{\YESS}{\ding{51}}
\newcommand{\NOO}{\ding{55}} %\textcolor{Red4}{\ding{55}}

\begin{table}[!b]
\caption{
    Number of packages with and without custom configuration scripts in Alpine Linux main and community repositories.
    Some packages (Safe=\NOO) contain scripts that break OS integrity.
}
\center
\begin{tabular}{m{1cm}cclm{3cm}l}
\firsthline
\multicolumn{4}{c}{Alpine repository} \\
\cline{1-4}
\multicolumn{2}{c}{\ftextnumero\xspace packages in} \\
\cline{1-2}
Main     & Community &  &  \\
\hline
5665 & 5916  & Total & Safe  \\
\hline
5531 & 5772  & Without scripts & \YESS \\
24  & 29    & With safe scripts & \YESS \\
110  & 115    & With unsafe scripts & \NOO \\

\lasthline
\end{tabular}
\label{tab:repo_stats}
\end{table}

\subsection{Solution to Problem 1: Sanitization}
\label{sec:design_sw_updates}
To enable support for software updates, we must solve two problems. 
First, convince a monitoring system that the integrity measurements of files extracted from the software package to the OS are valid. 
Second, make sure that the execution of a software package installation script does not cause the transition of the OS into an untrusted state.

%\textcolor{blue}{
  To address these problems, we introduce the concept of package sanitization (\autoref{fig:highlevel} (\raisebox{-1pt}{\ding{203}})).
  It consists of verifying and modifying packages by 
  i) changing installation scripts to ensure that their execution changes the OS configuration in a deterministic way; 
  ii) predicting such configuration;
  iii) including digital signatures of files delivered with the software package and the predicted OS configuration. 
%}

\myparagraph{Digital signatures}
Following the work of Berger et al. 2015, we propose that for each file stored inside a package, a corresponding digital signature certifying its integrity is also stored inside the package.
The package manager would extract digital signatures to the filesystem, allowing the IMA to include digital signatures inside the attestation report.
Consequently, the verifiers could recognize that the new integrity measurements are valid because they correspond to installation scripts and package-specific files.

\myparagraph{Installation scripts}
\label{sec:sanitization}
Software packages might contain scripts that are executed with administrative rights during the package installation.
Developers or package creators provide such scripts, and there are no limitations on what kind of OS configuration changes scripts can do.
Therefore it is possible that, due to a misconfiguration, a script reconfigures OS, allowing remote access to the machine.
We designed \sys to modify packages in such a way the installation scripts change OS configuration deterministically.
The packages which scripts cannot be sanitized are rejected from \sys, and thus not available for installation.

\newcommand{\YES}{\ding{51}}
\newcommand{\NO}{\ding{55}} %\textcolor{Red4}{\ding{55}}

\begin{table}[!tb]
\caption{
    Operations performed by installation scripts located in software packages in Alpine Linux repositories.
    Some operations (Safe=\NO) break OS integrity.
    The last column ("TSR") indicates which operations are safe after the sanitization.
    \emph{Filesystem changes} - add/remove/modify folders, symbolic links, and their permissions.
    \emph{Empty scripts} - conditional checks, display information.
}
\center
\begin{tabular}{cclcc}
\firsthline
\multicolumn{5}{c}{Operations executed in scripts} \\
\cline{1-5}
\multicolumn{2}{c}{\ftextnumero\xspace packages in} \\
\cline{1-2}
Main     & Community & Type & Safe & TSR \\
\hline
30 & 15  & Filesystem changes  & \YES & \YES  \\
5 & 17   & Empty scripts  & \YES & \YES  \\
17 & 19  & Text processing & \YES & \YES  \\
11  & 7  & Configuration change & \NO & \NO  \\
1  & 0   & Empty file creation & \NO & \YES  \\
97 & 104  & User/Group creation & \NO & \YES  \\  
4  & 6   & Shell activation & \NO & \NO  \\
\lasthline
\end{tabular}
\label{tab:repo_stats_all}
\end{table}

% \begin{tabular}{lccc}
% \textbf{Operation type} & 
% \textbf{Break OS integrity?} & 
% \textbf{Alpine Main } & 
% \textbf{Alpine Community (packages)} \\
% \hline
% Modify file system & no & 57 & 17 \\
% Text processing & no & 32 & 28 \\
% Modify OS configuration & & & \\
% output redirection to a file & yes & 8 & 3 \\
% create empty file & yes & 1 & 0 \\
% create new user  & yes & 87 & 79 \\  
% create new group & yes & 90 & 77 \\
% activate shell & yes & 4 & 6 \\
% \hline
% \end{tabular}

% % mv, mkdir, ln, chmod, rmdir, chown, cp, rm
% % findfs, nlplug-findfs, head, ls, hostname, stat
% % grep, false, awk, sed, expr, tr, printf, test, sort
% %  & cat, echo
% %  & touch
% %  & adduser 
% %  & addgroup
% %  & add-shell
% %  & passwd

To design the script sanitization algorithm, we started by analyzing existing scripts wrapped inside packages available in the Alpine Linux repositories\footnote{v3.11 of the Alpine Linux main \cite{alpine_repo_main} and community \cite{alpine_repo_community} repositories.}. 
\autoref{tab:repo_stats} shows that 97.6\% of packages do not contain any scripts. 
81\% of the remaining packages contain scripts that alter the OS configuration, breaking the system integrity. 

We analyzed commands executed inside the scripts to understand how they interfere with the OS configuration. 
\autoref{tab:repo_stats_all} shows that 45 packages modify the filesystem structure (\ie, copying, moving, or removing files, directories, and symbolic links, also changing their permissions).
From the OS integrity point of view, these actions are safe -- they do not violate system integrity as defined by the IMA. 
Similarly, 36 packages execute text processing utilities (\eg, parsing existing OS configuration), which do not alter any existing file; thus, they are safe.
% There are 22 packages which contain scripts which the only purpouse 
However, 230 packages contain scripts modifying the OS configuration, creating new users and groups, activating new shells, or creating empty files.
These scripts are unsafe because they modify existing file contents in which integrity is certified using pre-generated signatures (as discussed in the previous section).

\myparagraph{Script sanitization}
As we show next, the majority of the unsafe scripts provide a predictable output.
Hence it is possible to predict the OS configuration before installing the package.

The installation or update of 201 packages results in the creation of new users or groups.
In the case of Linux-based operating systems, three files are affected, \ie, /etc/passwd, /etc/group, /etc/shadow.
Interestingly, these files change in a deterministic way. 
Adding a new user or group results in adding a new well-defined line in at least one of these files.
However, the order in which users and groups are created determines final file contents.
In particular, different package installation order results in a different order in which users and groups are defined inside of each file.

Our solution consists of scanning the entire repository to learn about all possible users and groups that might be added by any software package.
Then, we change each installation script in each package in a way the script creates all possible users and groups in the same predefined order.
Consequently, any selection of packages and their order always results in the same OS configuration -- it contains all users and groups.
% Such approach is neutral form the security point of view because such created users can be used only to provide processes isolation.
Finally, \sys issues digital signatures over the predicted contents of the configuration files and modifies scripts to install the signatures in the target OS.
Monitoring systems accept the new OS configuration because they read a measurement report containing the signatures, which vouch for the new configuration files contents.

Our \sys implementation detected and sanitized two packages that not only create a user but also set an empty password and shell.
Installation of such packages  might cause a security breach by allowing an adversary to remotely connect to the OS using a well-known username and password \cite{CVE_2019_5021}.
We reported our findings to the Alpine Linux community.

% Although setting a new user for different processes is considered as a good practice, setting a user with a password and a valid shell might cause a security breach by allowing an adversary to remotely connect to the OS.

% We created CVEs and reported the packages to the Alpine community.

\myparagraph{Unsupported scripts}
\sys does not support 28 packages (0.24\%) out of all packages available in Alpine repositories.
%\textcolor{blue}{
  In particular, \sys does not support packages in which installation changes arbitrary configuration files.
  For example, a package \textit{roundcubemail} is not supported because it generates an unpredictable configuration file containing a random session key.
  Although \sys could support it by generating the session key during the sanitization, such a solution would contradict the script functionality that provides a unique key per OS.
%}

%\textcolor{blue}{
On the other hand, \sys intentionally 
%}
does not support software packages providing different shells (\eg, mksh, bash, tcsh).
Their scripts modify the OS configuration by activating a newly installed shell using \emph{add-shell} command. 
Although \sys might use the same technique as with adding users and groups, we argue that the installation of a custom shell should not occur during an OS update but should instead be part of the initial OS configuration. 

%  or should be rejected from \sys because they cause a security risk. 
% First, the integrity measurements of an empty file are always the same because one-way hash function produces a deterministic digest.
% Therefore, a monitoring system can include such values inside the whitelist.
% Another 30 packages created a user with a configured shell. 
% \sys supports packages that require a custom user and group, eliminating the security risks.

% `echo`: main/mkinitfs, main/cvechecker, community/roundcubemail, community/alpine-desktop
% write command output to conf file (main/gtk, main/cjdns)
% writes startup init.d script (main/alpine-baselayout)

% packages that set user and password: community/gitea, community/gogs

\subsection{Solution to Problem 2: Proxy}
\label{sec:design_community}
We designed \sys as a proxy between package managers and software repositories provided by the community.
This design decision permits \sys to act as a separate software repository that serves sanitized packages signed directly by \sys.
From the community point of view, no changes are required to the existing software package creation processes, software package formats, or the implementation of package managers. 
Package managers recognize \sys as a standard repository mirror.
Hence, it is enough to adjust the OS configuration in a way the package manager uses only \sys as a mirror.

% Then, \sys splits the package into three archives and checks the authenticity and integrity of the package by verifying the signature using the signer public keys defined in the policy. 
% Then, \sys reads the integrity digest from the metafile and verifies the integrity of the third archive. 

\subsection{Solution to Problem 3: Shielded execution}
\label{sec:design_key_mgmt}
\sys requires a signing key to certify changes made to packages during the sanitization process.
To protect the signing key from an adversary with root access to the machine, we propose to use \gls{tee}.
In particular, we propose to leverage \gls{sgx}, which is Intel's \gls{cpu} extension providing confidentiality and integrity guarantees to applications running in environments in which OS, hypervisor, or \gls{bios} might have been compromised.
Other studies \cite{palaemon_2020} demonstrated that applications running inside an enclave (a trusted execution environment provided by \gls{sgx}) can generate, store, and use cryptographic keys that are only known to the specific application -- not even a human being can read them.
\sys's design relies on that concept.
By running \sys inside an enclave, \sys generates a signing key that is used later to sign all modified software packages.
The public portion of the signing key is exposed to both operating systems and monitoring systems that use it to verify that software packages were created by \sys.
% , such as asymmetric keys \cite{rsa_1978, koblitz_elliptic_1987} required to issue a software package digital signature.

% TODO: wrapp signing key / share it with other \sys instances
% To support failures, we propose to 
% In case of a downtime, a repository can be recreated on another computer by deploying the same policy.

\lstdefinestyle{interfaces}{
  float=tbp!,
  floatplacement=tbp,
  abovecaptionskip=-40pt,
}
\begin{lstlisting}[caption=Policy example,label={lst:policy},language=yaml,breaklines=true,breakatwhitespace=true,escapechar=^]
mirrors: ^\label{policy:mirrors_start}^
  - hostname: https://alpinelinux/v3.10/
    certificate_chain: |- ^\label{policy:mirror_cert_chain}^
      -----BEGIN CERTIFICATE-----
      (...)
      -----END CERTIFICATE-----
  - hostname: https://yandex.ru/alpine/v3.10/
    certificate_chain: |- 
      -----BEGIN CERTIFICATE-----
      (...)
      -----END CERTIFICATE-----
  - hostname: https://ustc.edu.cn/alpine/v3.10/
    certificate_chain: |- 
      -----BEGIN CERTIFICATE-----
      (...)
      -----END CERTIFICATE----- ^\label{policy:mirrors_end}^
signers_keys: ^\label{policy:signers_keys_start}^ 
  - |- # e.g., alpine@alpinelinux.org-4a40.rsa.pub
    -----BEGIN PUBLIC KEY-----
    (...)
    -----END PUBLIC KEY-----
  - |- # e.g., alpine@alpinelinux.org-524b.rsa.pub
    -----BEGIN PUBLIC KEY-----
    (...)
    -----END PUBLIC KEY----- ^\label{policy:signers_keys_end}^
init_config_files:
  - path: /etc/passwd
    content: |-
      root:x:0:0:root:/root:/bin/ash
      daemon:x:2:2:daemon:/sbin:/sbin/nologin
      (...)
  - path: /etc/shadow
    content: |-
      root:$6$UmJDHY...25/:18206:0:::::
      daemon:!::0:::::
      (...)
  - path: /etc/group
    content: |-
      root:x:0:root
      daemon:x:2:root,bin,daemon
      (...)
\end{lstlisting}
%\textcolor{blue}{added: init\_config\_files section inside the policy Listing 1 (above)}

\subsection{Solution to Problem 4: Quorum}
\label{sec:byzantine_mirrors}
An adversary might leverage administrative privileges to drop network traffic to certain hosts.
In particular, she might prevent \sys from accessing the original repository, forcing \sys to rely on a mirror serving outdated software packages.
%\textcolor{red}{removed: redundant information (provided in the subsection title) about the problem it refers to}

As specified in \S\ref{sec:threatmodel}, we assume that the majority of repository mirrors are available and provide the latest snapshot of the original repository.
\sys does not trust any individual mirror.
Instead, it reads \emph{2f+1} mirrors and only relies on the information that matches responses of at least \emph{f+1} mirrors.
Importantly, \sys requires a quorum only when reading the metadata index. 
The packages can be downloaded from a single mirror because their integrity is verifiable using the metadata index.
% This assumption is valid because by definition the mirrors contain the exact copy of the main software repository.

To allow different organizations to specify individual security requirements (\ie, which mirrors to use, which package creators to trust) 
and to provide custom initial OS configuration (\ie, initial users, groups, and passwords), 
%\textcolor{blue}{and to provide custom initial OS configuration (\ie, initial users, groups, and passwords), }
\sys accepts security policies.
\autoref{lst:policy} shows an example of such a security policy.
The format permits defining a list of mirrors (lines \ref{policy:mirrors_start}-\ref{policy:mirrors_end}) and a list of trusted package signers (lines \ref{policy:signers_keys_start}-\ref{policy:signers_keys_end}). 
The package signer is a developer or a build system (\eg, \glsdesc{cicd}) that builds, signs, and deploys packages to the original repository. 

\sys enforces the security policy by publishing only software packages in versions offered by the majority of available mirrors and only created by trusted entities. 
%\textcolor{blue}{
  The policy could be extended to support a private/closed variant in which an OS owner can specify a subset of supported software packages by specifying whitelist/blacklist of packages.
%}

\begin{figure}[tbp!]
\centering
% \captionsetup{skip=4pt}
\includegraphics[width=0.48\textwidth]{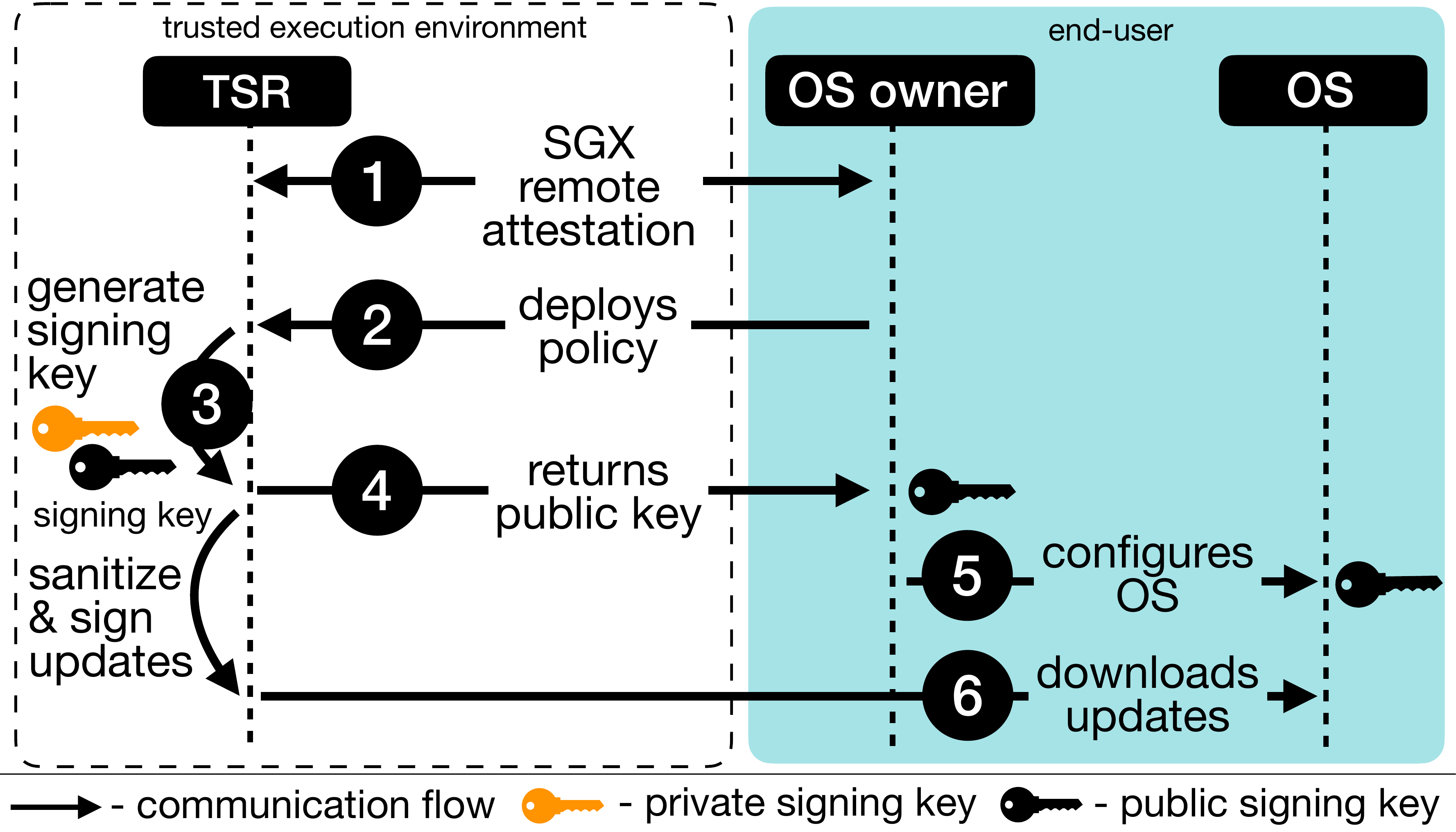}
\caption{
    The protocol of distributing the public portion of the signing key, which can be used to verify the authenticity of the software packages.
}
\label{fig:key_distribution}
\end{figure}

\autoref{fig:key_distribution} shows how an organization can deploy a security policy to \sys.
First, it establishes trust with \sys (\raisebox{-1pt}{\ding{202}}) using \gls{sgx} remote attestation protocol \cite{johnson2016intel}, which permits ensuring that \sys executes inside an enclave on the genuine Intel \gls{cpu}.
Then, it uploads the security policy (\raisebox{-1pt}{\ding{203}}), causing \sys to generate a new signing key (\raisebox{-1pt}{\ding{204}}), to store the security policy, and to return the public portion of the newly generated signing key (\raisebox{-1pt}{\ding{205}}).
Finally, the public key is distributed to all integrity-enforced operating systems and integrity monitoring systems (\raisebox{-1pt}{\ding{206}}).
At this point, the OS accepts sanitized software packages (\raisebox{-1pt}{\ding{207}}), and the integrity monitoring system accepts integrity measurements of files digitally signed by \sys.
%\textcolor{blue}{
  In more detail, the integration between integrity monitoring systems and \sys consists of adjusting integrity monitoring systems configuration to trust \sys signing key. Hence, integrity monitoring systems accept integrity measurements signed by \sys. 
  \sys returns the signing key during the repository initialization (\S\ref{sec:repo_init}) triggered by the OS owner (\autoref{fig:key_distribution}).
%}

%!TEX root = paper.tex
\section{Implementation}
\label{sec:implementation}
We developed \sys in Rust, a programming language that ensures memory safety~\cite{matsakis_rust_2014}. 
We rely on the external Rust libraries, \ie, Hyper \cite{hyperRs}, Rustls ~\cite{rustls}, to build the \gls{rest} \gls{api}~\cite{fielding_information_2000}. 
We use a Rust-based crypto library ring~\cite{rust_ring} to issue digital signatures. 
%\textcolor{red}{removed: 'Furthermore'}
We use SCONE Rust cross-compilers \cite{sconecuratedimages_rust} to execute \sys inside an SGX enclave. 
\sys is about 3.3k \glsdesc{sloc}, excluding external libraries. 

%\textcolor{blue}{
    We rely on SGX because it provides the following properties: \emph{confidentiality} to protect the signing keys, 
    \emph{integrity} to protect the sanitization process, and \emph{attestation protocol} to remotely ensure TSR integrity during the policy deployment. 
    Alternative TEEs~\cite{keystone_tee, amd_sev_api, intel_txt_whitepaper, flicker2008} providing similar functionality might be considered but
    the threat model should be carefully adjusted, according to TEE-specific implementation. 
    For example, TEEs relying on late-launch technologies~\cite{amd_sev_api, intel_txt_whitepaper, flicker2008} must assume trusted link between CPU and TPM ~\cite{winter_hijackers_2013, winter_hijackers_bus_2012}, while others, like Keystone~\cite{keystone_tee}, must assume trusted boot process. 
%}

% The external libraries and their dependencies add extra XX LOC.

\subsection{Supported package formats}
Our prototype implementation of \sys supports \emph{apk} packages used by Alpine Linux.
We selected Alpine Linux because it is a popular security-oriented Linux distribution that minimizes the amount of software required to run the OS. 
It is an important property for systems relying on trusted computing. 
% In particular, Alpine Linux is characterized by relatively small \gls{tcb}, which plays an important role in systems relying on trusted computing. 
In the future, we plan to add support for other formats (\ie, deb, rpm) used by other Linux distributions. 

% At this point, \sys analyzes the configuration scripts. 
% In case \sys detects the configuration script might break the OS integrity, \sys returns an error, \ie 404 HTTP response code. 

\subsection{Repository initialization}
\label{sec:repo_init}
\sys can be executed in the cloud and is operated by a cloud provider, who is responsible for correct hardware initialization, installation of the operating system, and \sys execution.
The cloud provider exposes the hostname on which \sys \gls{api} is accessible by his clients.

Multiple clients share a single \sys instance.
Each client deploys a policy to create his individual, logically separated, software repository within the \sys instance.
For each new repository, \sys, which runs inside an SGX enclave, generates a unique repository identifier and a unique signing key.
The identifier and the public portion of the signing key are returned to the client as a response to the policy deployment request issued via https.
Each client accesses his repository via the \gls{rest} \gls{api} after providing the identifier.
By verifying the digital signature of the package, the client ensures that the package conforms to his requirements defined inside the policy.

\subsection{Package sanitization}
We define package sanitization as an operation consisting of the following steps: verifying package integrity and authenticity, extracting files from the package archive, modifying the installation scripts (see \S\ref{sec:sanitization}), issuing digital signatures to all files inside the package, updating the metafile, and recreating the package.
\sys issues digital signatures using the signing key generated during the policy deployment. 

The digital signatures are stored inside \gls{pax} headers \cite{pax_headers_format} of the tar archive \cite{tar_format}, which is logically equivalent to the package.
The modern versions of tar extractors (\eg, GNU tar~\cite{gnu_tar}) transparently copy the specific PAX headers' value into the extended attributes in the filesystem. 
Before opening a file, Linux IMA scans extended attributes and includes the digital signature inside a dedicated file (IMA log). 
Consequently, the monitoring systems read the measurement report and the IMA log.
They check the integrity of every file measured by the IMA by verifying its digital signature included inside the IMA log.
% To get independent of the tar format, an alternative solution using mtree \cite{mtree_format} might be used.

\subsection{OS configuration}
Software repositories include information about software packages sizes and hashes inside the repository metadata index to mitigate the endless data attack and the extraneous dependencies attack \cite{cappos_look_2008}.
Operating systems read the package size and its hash from the metadata index to ensure they download the file of the expected size and contents.
Because of that, when an OS requests \sys to return the metadata index for the first time, \sys downloads and sanitizes all packages listed in the upstream metadata index.
Then, \sys generates a new metadata index that matches the sanitized packages and returns it.
Although the first metadata index generation is time-consuming, subsequent requests require \sys to sanitize only packages that have changed on the upstream mirrors, since the previous read.

Each integrity-enforced OS must be reconfigured to use the \sys repository instead of mirrors.
Moreover, the OS must trust the packages signed by \sys; thus, the public portion of the signing key must be added to the list of trusted signers.
This reconfiguration can be done automatically using configuration management systems such as Puppet \cite{puppet} or Chef \cite{chef_io}.

\subsection{Package caching}
A slow read of software updates increases the vulnerability window for the \gls{toctou} attack, where an adversary exploits the existing vulnerabilities until the security patches become available in the repository. 
In the case of \sys, this time is increased by the sanitization process (see \S\ref{sec:sanitization}) and the time required to read the majority of available mirrors (see \S\ref{sec:byzantine_mirrors}).

To minimize the vulnerability window for the \gls{toctou} attack, \sys uses a local file system to cache the already sanitized packages, including the metadata index.
\sys detects the outdated software packages each time \sys reads the new metadata index from the upstream mirrors.
Consequently, \sys invalidates the metadata index, downloads the new version of the package, sanitizes it, and stores the new version inside the cache.

An adversary might tamper with the cache by reverting software packages and the metadata index to the outdated versions.
To mitigate the attack, \sys stores metadata indexes (the latest one read from upstream mirrors and the one reflecting the already sanitized packages) inside its memory, which integrity and freshness are guaranteed by \gls{sgx}.
\sys uses the first metadata index to check which software packages changed in the upstream mirrors.
It uses the second metadata index to verify that the package read from the cache (untrusted disk) has not been rollbacked, before returning it to the OS.

However, the data stored inside \sys memory is lost as soon as \sys is shutdown, for example, due to the OS restart.
To preserve the metadata indexes across \sys restarts, we extended \sys implementation with support for \gls{tpm} \gls{mc} \cite{tpm_2_0_spec}.
After generating the metafile, \sys increases the \gls{mc} value and uses SGX sealing \cite{anati2013innovative} to store the metadata indexes together with the MC value on the disk.
The SGX sealing, and its revert operation unsealing, uses a CPU- and enclave-specific key. 
Hence, only the same enclave running on the same CPU can unseal the previously sealed file.
After the restart, \sys unseals the metadata indexes from the disk together with the MC value and verifies that the unsealed MC value matches the current MC value.

% The next level of caching can be provided by placing load balancers in front of the \sys instance.
% Each load balancer might cache the packages, reducing the amount of requests hitting the \sys instance.

%!TEX root = paper.tex
\newcommand{\ubuntuver}{Ubuntu 18.04\xspace}
\newcommand{\alpinever}{Alpine 3.10\xspace}
\newcommand{\osver}{Alpine 3.10\xspace}

\newcommand{\kernelver}{Linux kernel 4.19.58-vanilla\xspace}
\newcommand{\machinetype}{Dell PowerEdge R330\xspace}
\newcommand{\machinecpu}{Intel Xeon E3-1280 v6 CPU\xspace}
\newcommand{\tpmchipversion}{Infineon 9665 TPM 2.0\xspace}
\newcommand{\microcodever}{0x5e\xspace}

\section{Evaluation}
\label{sec:evaluation}
In this section, we evaluate \sys to answer the following questions:
\begin{enumerate}[label=\emph{\textbullet}]
\item What is the overhead related to the package sanitization?
\item What are the performance limitations incurred by running \sys inside an SGX enclave?
\item What is the cost of tolerating compromised mirrors?
\end{enumerate}

\myparagraph{Testbed.}
Experiments execute on a rack-based cluster of \machinetype servers equipped with an \machinecpu, 64 GiB of RAM, Samsung SSD 850 EVO 1TB.
All machines have a 10 Gb Ethernet \gls{nic} connected to a 20 Gb/s switched network. 
The support for \gls{sgx} is turned on; the hyper-threading is switched off. 
We statically configured \gls{sgx} to reserve 128 \unit{MB} of RAM for the \gls{epc} \cite{costan2016intel}. 
The \glspl{cpu} are on the microcode patch level \microcodever.
We run Alpine Linux 3.10 with enabled Linux IMA.

\subsection{Package sanitization overhead}
\label{eval:sanitization}
The sanitization process directly influences the software update process, \ie, time after which software updates are visible by the OS and the latency taken by the OS to download the update.
For that reason, we run experiments in which we instrumented the sanitization process to measure its impact on packages from the main and community repositories of Alpine Linux.
The results are based on a 20\% trimmed mean from six independent experiment executions.

\begin{shaded*}
\begin{quoting}
\centering
How much time does it take to sanitize all packages?
\end{quoting}
\end{shaded*}

\begin{table}[!t]
\caption{
    Time required to initialize a repository.
    We assume two scenarios. 
    In the optimistic one, \sys has access to a copy of packages stored in a cache.
    In the pessimistic one, during the policy deployment, \sys must download all packages from the original repository. 
}
\center
\begin{tabular}{lll}
\firsthline
\multicolumn{2}{c}{Time} \\
\cline{1-2}
pessimistic & optimistic & Operation \\
\hline
17 min  & 0 min & Download packages  \\
< 1 min & < 1 min & Policy deployment \\
13 min  & 13 min & Sanitize packages  \\
\hline            
30 min  & 13 min & Total \\
\lasthline
\end{tabular}
\label{tab:repo_init}
\end{table}
From the OS perspective, low repository initialization time results in faster delivery of software updates.
Therefore, we calculated the time requires to create a new repository, \ie, to download and to sanitize all packages.
In the case of packages update, this time is expected to be significantly lower because \sys would have to download and to sanitize just a small amount of packages.

% \definecolor{green}{rgb}{0.1,0.1,0.1}
% \newcommand{\done}{\cellcolor{teal}done}  %{0.9}
% \newcommand{\hcyan}[1]{{\color{teal} #1}}

\begin{table}[!b]
\caption{
    Spearman rank correlation coefficients ($\rho$) relating the package-specific properties and sanitization-specific operations. 
    The corresponding p values are indicated by regular font in grey fields (p < 0.05), 
    bold font in grey fields (p < 0.001);
    % as well as italic and underlined font in grey fields (p < 0.001); 
    fields with regular font indicate p > 0.05.
}
\center
\begin{tabular}{lcc}
\firsthline
        & number of files & package size\\
\hline
archive, compress        & \cellcolor{gray!20}.46 & \cellcolor{gray!20}\textbf{.61} \\
check integrity       & \cellcolor{gray!20}\textbf{- .62} & \cellcolor{gray!20}\textbf{- .93} \\
generate signatures   & \cellcolor{gray!20}\textbf{.69} & \cellcolor{gray!20}\textbf{.03} \\
modify scripts        & -.27 & \cellcolor{gray!20}\textbf{- .33} \\
\lasthline
\end{tabular}
\label{tab:correlations}
\end{table}

% he corresponding p values are indicated by regular font in grey fields (p < 0.05), 
% bold font in grey fields (p < 0.01), 
% as well as italic and underlined font in grey fields (p < 0.001); 
% fields without grey shading indicate p > 0.05.

\autoref{tab:repo_init} shows the time taken to establish a new repository, assuming two scenarios.
In the optimistic scenario, which takes about 13 min, \sys has access to pre-fetched packages, which are available, for example, pre-fetched by a service provider.
In the pessimistic one, which takes about 30 min, \sys additionally downloads original packages (about 3 GB of data) from upstream repositories.
We argue that the download time can be greatly reduced by enabling parallel downloading.
This performance improvement is left as part of future work. 

\begin{shaded*}
\begin{quoting}
\centering
What are the main factors driving the sanitization time?
\end{quoting}
\end{shaded*}

\sys sanitizes all packages provided with a software update, thus introducing a delay in how fast the OS receives the update.
Therefore, it is important to understand the main drivers controlling the sanitization time.

\autoref{tab:correlations} shows the correlations between package-specific properties (\ie, number of files inside a package, package size) and the proportional time contribution of certain components of the sanitization time.
%  (\ie, package de/compression, package authenticity and integrity verification, generation of digital signatures, modifications of installation scripts).
We observe a strong positive correlation ($\rho$ = 0.61) between the archive processing time and package size, which indicates that the archive, compression and decompression algorithms take more time to process bigger archives.
Also, we observe a strong correlation ($\rho$ = 0.69) between signatures generation and the number of files inside a package.
It confirms the intuitive expectation that in packages containing many files, the signature generation becomes a dominant factor of the sanitization time. 
Furthermore, we explain that a strong negative correlation ($\rho$ = -0.93) between checking the package integrity and package size shows that the time required to check the package integrity becomes negligible for bigger packages because other operations (\ie, signature generation, archive, compression and decompression) become the dominant factors.
All in all, we anticipate that the sanitization time is mainly driven by
1) extracting files from a package and compressing them again into a package, 
2) issuing digital signatures.

\begin{figure}[tbp!]
  \centering
  \includegraphics[width=0.45\textwidth]{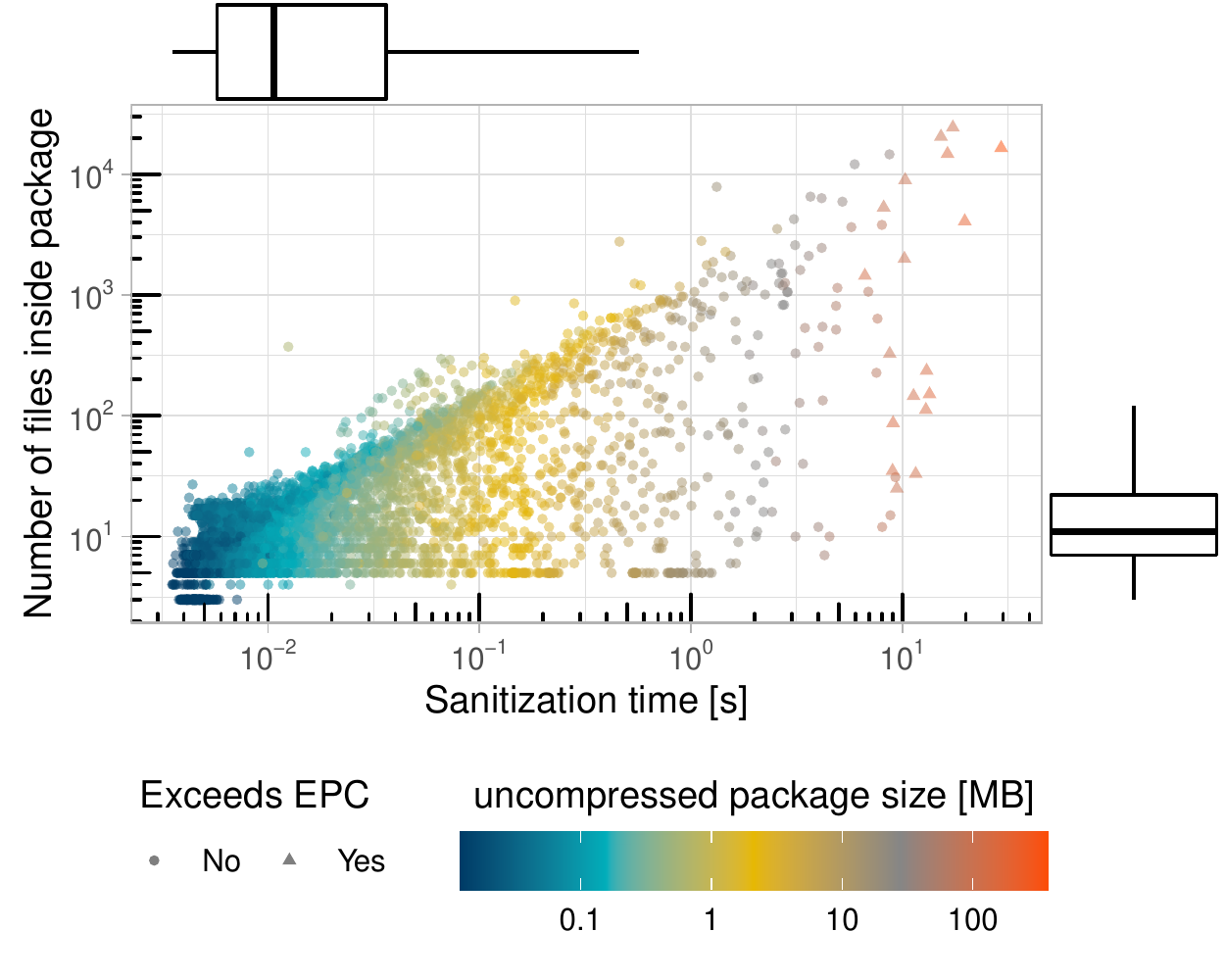}
  \vspace{-2mm}
  \caption{
	Time required to sanitize a package, depending on the number of files and size.
  Color represents package size after decompression.
  Packages which size exceeds the EPC are marked as $\blacktriangle$.
  Boxplots indicate 5th, 25th, 50th, 75th, and 95th percentile.
  }
  \label{fig:load_times}
\end{figure}

% \autoref{fig:load_times} shows that the sanitization time increases with the package size and the number of files inside the package.
% The sizes of 95\% of packages do not exceed 5.8 MB, and their processing time is less than 422 ms.
% Therefore,  the sanitization time should not 

% It might indicate that the high packages sanitization times are caused by the operations sensitive to the package size, \ie, extracting files from an archive, .
%  sanitization time is mostly driven by the package extraction and less by the creating signature  package size is a dominant factor.

% It takes an average of 143 ms to sanitize a single package.

\begin{shaded*}
\begin{quoting}
\centering
How much time does it take to sanitize a package?
\end{quoting}
\end{shaded*}

To better estimate time which \sys requires to expose an update, we examine the time it takes to sanitize individual packages.
\autoref{fig:load_times} shows the relationship between sanitization time and package-specific properties, such as the package size and the number of files inside the package.
The sanitization time is not evenly distributed; it changes from 11 ms (50th percentile), 36 ms (75th percentile), 422 ms (95th percentile), to 30 seconds (100th percentile).
% We observe that the sanitization time increases with the number of files inside the package and with the package size.

\begin{shaded*}
\begin{quoting}
\centering
What is the impact of sanitization on the repository size?
\end{quoting}
\end{shaded*}

Repository size is the sum of all packages served by the repository.
The higher the size, the more resources (\ie, disk space, bandwidth) are utilized.
It not only increases the maintenance costs but also increases the latency because the OS requires more time to download packages.

\begin{figure}[tbp!]
  \centering
  \includegraphics[width=0.45\textwidth]{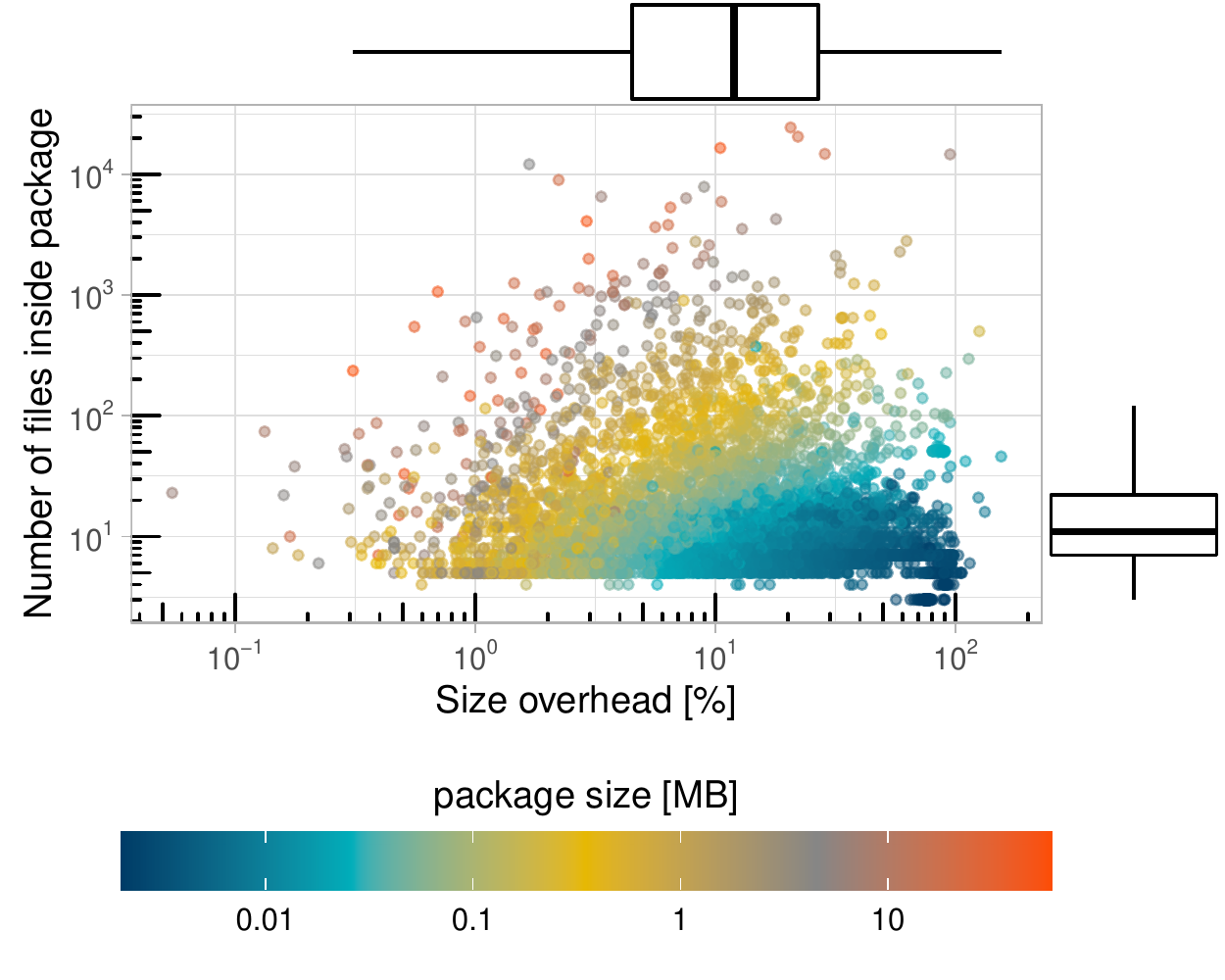}
  \vspace{-2mm}
  \caption{
	  Increase of package size caused by sanitization, depending on the number of files inside the package.
    Color represents size of a package (files are compressed into a single archive).
    Boxplots indicate 5th, 25th, 50th, 75th, and 95th percentile.
  }
  \label{fig:size_overhead}
\end{figure}

\autoref{fig:size_overhead} shows that the package sizes increase when compared to the original package size and the number of files located inside the package.
In particular, the sanitization process increases package size by 12\%, 27\%, and 76\% in 50th, 75th, and 95th percentile, respectively.
Packages with many small files suffer most from sanitization because the sizes of file signatures (each signature is 256 bytes) constitute a dominant part of the total package size.
However, the total repository size increases only by 3.6\%, from 3000 MB to 3110 MB.

% The sanitization process increases the package size because 
% 1) it adds one digital signature for one file located inside the package, and 
% 2) it extends every installation scripts in every package with list of commands modifying OS configuration read from all packages.
% In this experiment, we measure how much 

% 50th size [MB]: 11.94
% 75th size [MB]: 26.83
% 95th size [MB]: 76.39

\begin{shaded*}
\begin{quoting}
\centering
Does the caching decreases the latency of package download?
\end{quoting}
\end{shaded*}

\sys implements caching to decrease the latency of accessing sanitized packages; it stores on the disk the original version of the package (the one fetched from upstream and not yet sanitized) and the sanitized one.
We run an experiment in which we measured how much time does \sys require to respond to a download request, assuming three scenarios:
\begin{enumerate*}[label=\emph{(\roman*)}]
    \item only the original packages are cached, (\emph{Original}),
    \item both original and sanitized packages are cached (\emph{Sanitized}), and
    \item packages are not available in the cache (\emph{None}).
\end{enumerate*} 

In the first scenario, \sys downloads packages from an official Alpine mirror located on the same continent (an average network latency 26.4 ms).
In the last two scenarios, \sys reads packages from the local disk.
In each scenario, we requested \sys to return every package available in the upstream Alpine repository sequentially.
We calculated the latency of downloading each package as a 20\% trimmed average from five repeated downloads.

% cache sanitized: 4.002607 ms
% cache original: 208.2737 ms
% no cache: 554.725 ms

\begin{figure}[tbp!]
  \centering
  \includegraphics[width=0.45\textwidth]{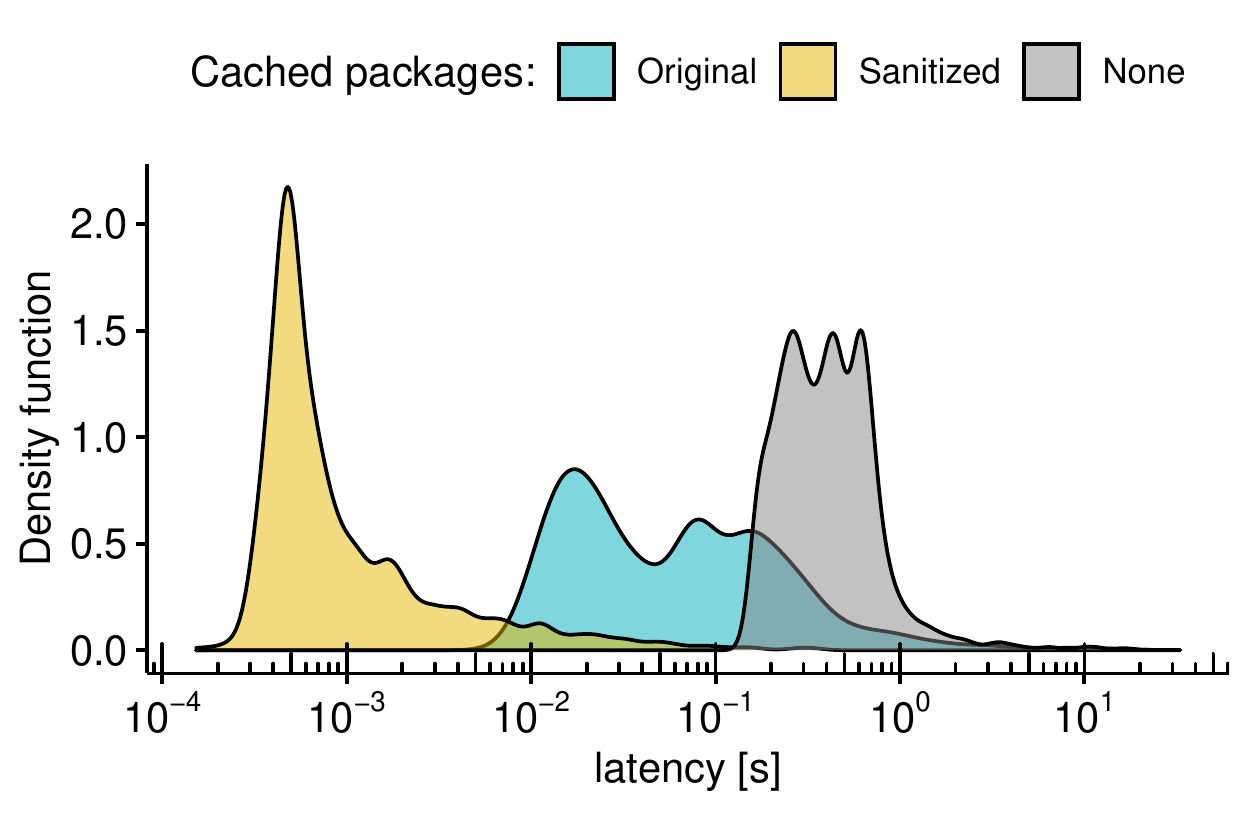}
  \vspace{-2mm}
  \caption{
    Comparison of package download latencies for scenarios in which \sys has access to original packages in the cache (\emph{Original}), has access to already sanitized packages (\emph{Sanitized}), and does not have access to any cached packages (\emph{None}).
  }
  \label{fig:caching}
\end{figure}

\autoref{fig:caching} shows distributions of package download latencies for the scenarios mentioned above.
Caching the sanitization results decreases the average download latency $129\times$ when compared to the scenario where \sys runs without cache.
We anticipate that the latency variation (0.37 ms) is mainly caused by accessing the cache (\ie, reading packages of different sizes) and verifying packages integrity after reading them from untrusted storage.

Similarly, caching the original packages decreases the average download latency $2.7\times$ when compared to the scenario where \sys runs without cache.
This is mostly the result of faster read of a package from the local disk than from a remote mirror accessed by the network.

% Because downloading packages from the original repositories and sanitizing them is a time-consuming process, \sys implements caching by storing both the original packages and the sanitized ones on disk.
% However, there are situations like getting free space on disk or cache invalidation when \sys cannot rely on the cached value. 
% Therefore, we measured how does the latency of accessing sanitized packages change when there is

\begin{shaded*}
\begin{quoting}
\centering
    %\textcolor{blue}{What is the end-to-end latency of installing an update sanitized by \sys?}
    What is the end-to-end latency of installing an update sanitized by \sys?
\end{quoting}
\end{shaded*}

%{ \color{blue} 
    Installation of a software update takes a considerable amount of time because a package manager must download and verify the update, prepare the system for the new package version (check dependencies, lock installed packages database), unpack the new software package, launch installation scripts, copy files, set permissions, and finally clean the filesystem from no longer necessary files.
    In this experiment, we check the end-to-end latency of installing an update, which consists of sanitized packages or native Alpine packages.
    We measure the update installation latency for more than 5000 packages cached in a repository, \ie, \sys serves sanitized packages from the cache.
    Before launching the experiment for each single package, we install the package, and then we tamper with the OS configuration to pretend the installed package is outdated.
    We do it by modifying the package version number and its integrity hash stored in the file-based database used by the Alpine Linux to store information about installed packages.
    Before measuring the next package, we uninstall the previously measured package from the OS.
    
    \begin{figure}[tp!]
    \centering
    \includegraphics[width=0.45\textwidth]{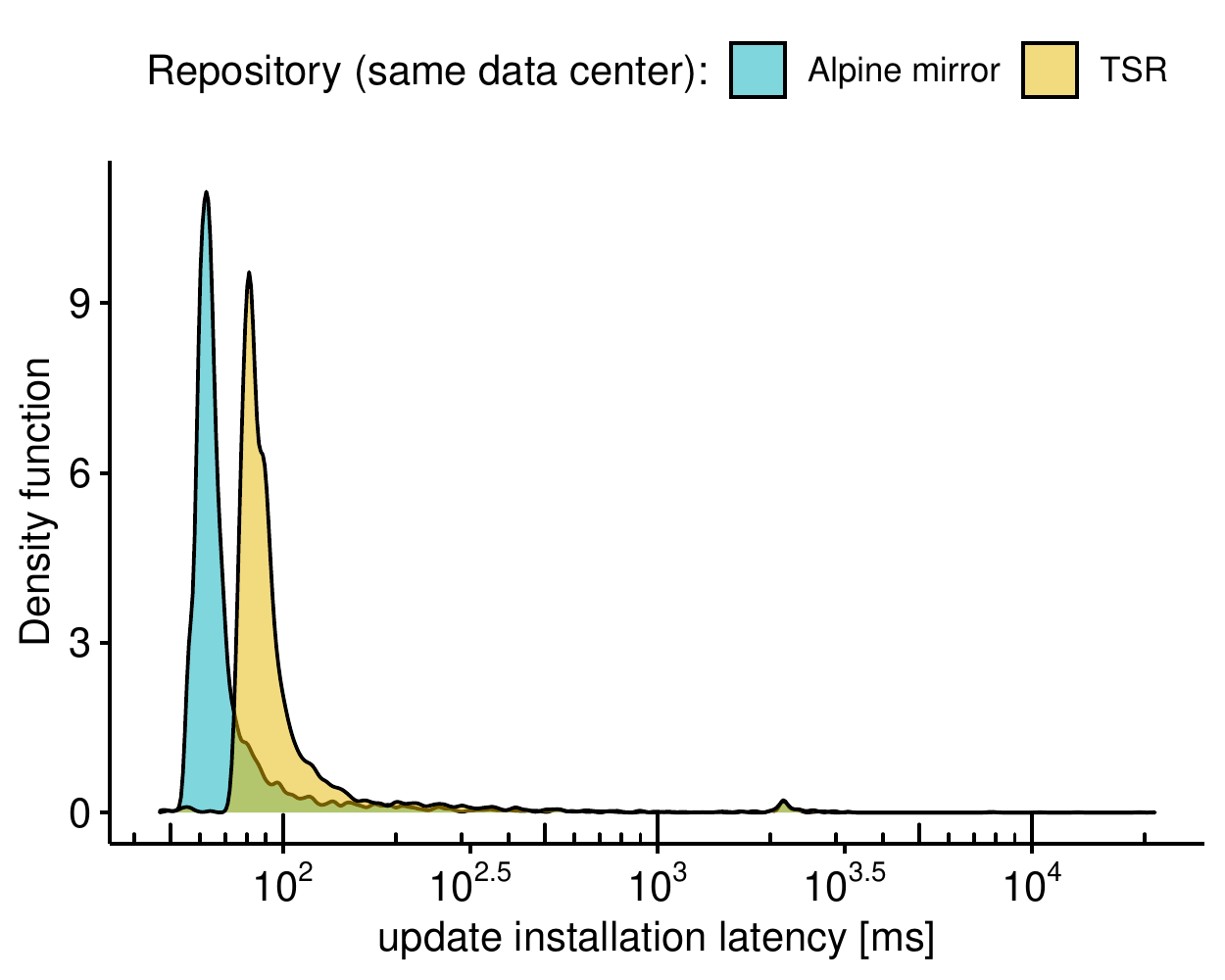}
    \vspace{-2mm}
    \caption{
        %\textcolor{blue}{End-to-end latency of installing software updates.}
        End-to-end latency of installing software updates.
    }
    \label{fig:endtoendtime}
\end{figure}

%  provided by \sys and an Alpine mirror located in the same data center.

    \autoref{fig:endtoendtime} shows the experiment results in which we use two repositories, \sys and Alpine mirror, located in the same data center. 
    We assume differences between network latency in both setups to be negligible.
    An average update installation latency is 141 ms and 110 ms for \sys and Alpine mirror, respectively.
    The higher latency observed when installing sanitized packages is caused by installing digital signatures in the filesystem.
%}

\subsection{SGX limitations}
The current version of the SGX has a limited memory, up to 128 \unit{MB} for SGXv1.
Applications exceeding this amount cause SGX to swap the memory leading to performance degradation.
Hence, we address the question of:

\begin{shaded*}
\begin{quoting}
\centering
What is the performance overhead of running \sys inside an SGX enclave?
\end{quoting}
\end{shaded*}

To answer this question, we observe that the package sanitization is the most memory consuming operation because \sys extracts and manipulates the package completely in the memory.
For that reason, we executed \sys without SGX to measure the processing time of all available packages. 

\begin{figure}[tbp!]
  \centering
  \includegraphics[width=0.45\textwidth]{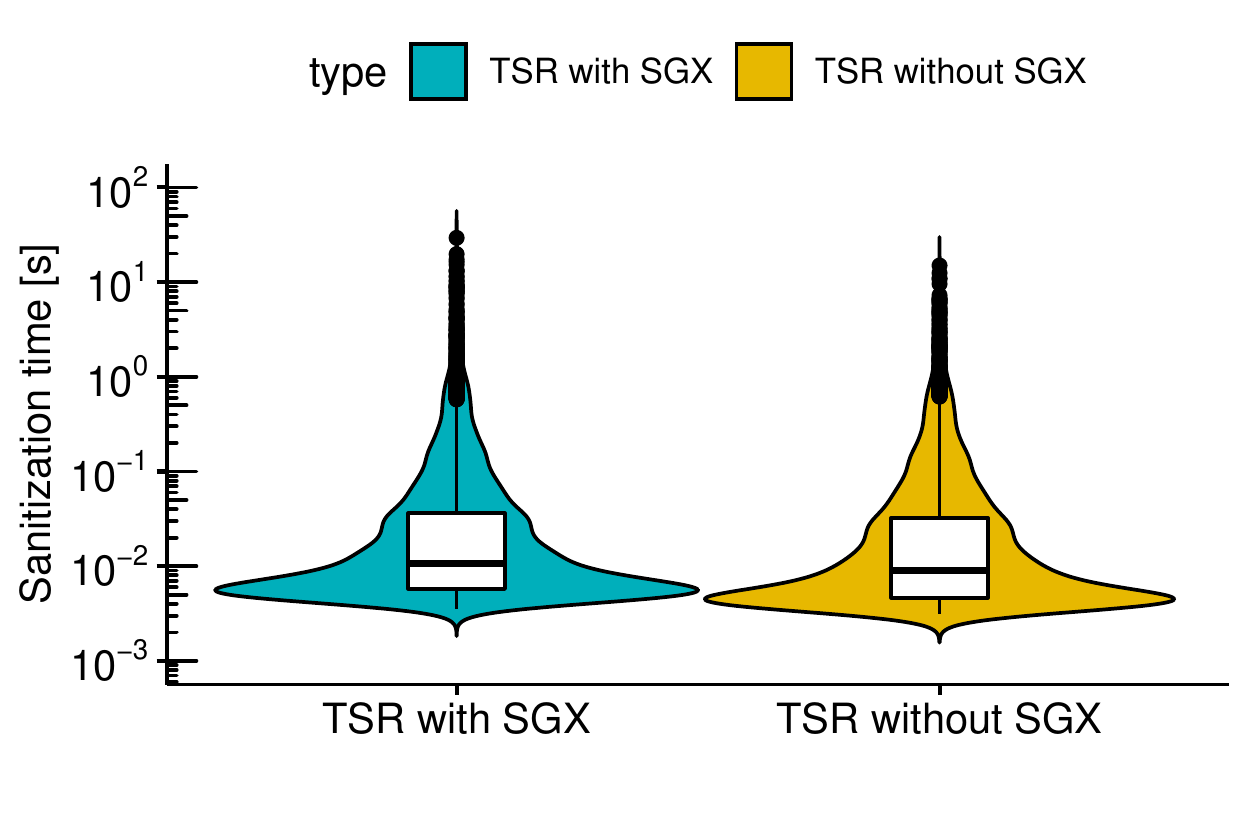}
  \vspace{-2mm}
  \caption{
	  Violin plot showing comparison of sanitization times executed inside and outside of an SGX enclave.
    Boxplots indicate 5th, 25th, 50th, 75th, and 95th percentile.
  }
  \label{fig:sanitization_no_sgx}
\end{figure}

\autoref{fig:sanitization_no_sgx} shows the comparison of packages sanitization times executed inside and outside an SGX enclave.
We observe a minor overhead of executing inside SGX; $1.18\times$ at 50th percentile, $1.12\times$ at 75th percentile, and $1.16\times$ at 95th percentile.
However, at the top 5 percentiles that represent packages with sizes exceeding EPC, the SGX overhead increases to $1.96\times$ because of EPC paging.
The total sanitization time required to process all packages in the repository increases from 9.5 min to 13.6 min ($1.43\times$) when running \sys inside an SGX enclave.

% total sanitization time (no sgx) = 570 s = 9.5 min
% total sanitization time (sgx) = 814 s = 13.6 min (1.43x times more than when without SGX)

% no sgx: 50th time: 9.06, 75th time: 32.45, 95th time: 365.31, Max time: 14.99, Mean time: 100.69
% sgx: 50th time: 10.69, 75th time: 36.22, 95th time: 422.53, Max time: 29.37, Mean time: 143.68

\subsection{Tolerating compromised mirrors}

\begin{shaded*}
\begin{quoting}
\centering
What is the overhead of mitigating compromised mirrors?
\end{quoting}
\end{shaded*}
In this experiment, we measured the latency in which \sys (running in Europe) returns the metadata index depending on the number of mirrors defined in the policy and their geographical locations. 
We were increasing the number of mirrors from one (default setting currently used by operating systems) to ten instances.
We divided the experiment into four scenarios.
In each scenario, \sys uses official Alpine mirrors located on different continents, \ie, Asia, Europe, North America, and their combination (\emph{All}).
In each scenario, we calculated a 10\% trimmed latency average from 20 consecutive requests.

\begin{figure}[bp!]
  \centering
  \includegraphics[width=0.45\textwidth]{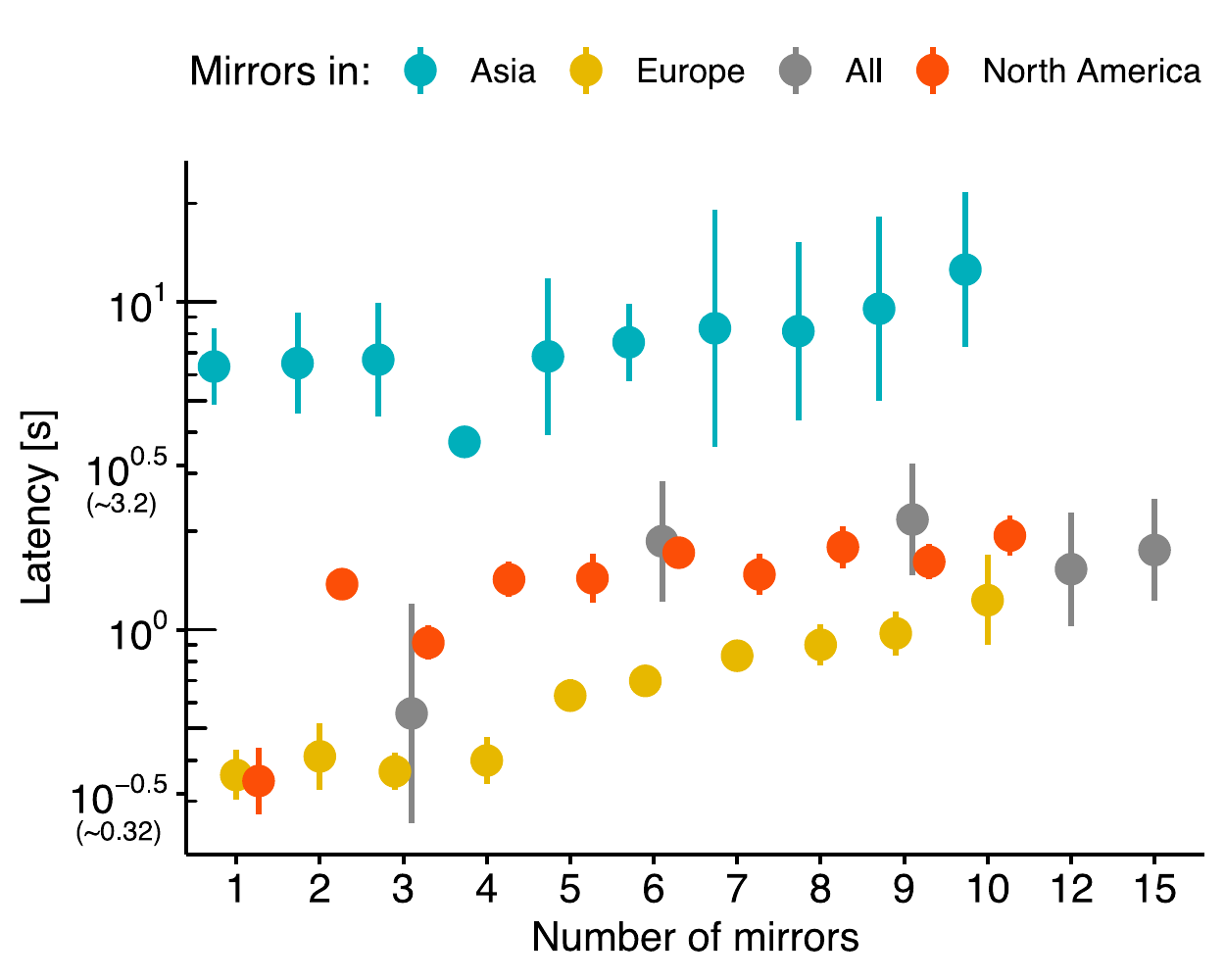}
  \vspace{-2mm}
  \caption{
      Latency of downloading the repository index from \sys.
      %, depending on the number and location of mirrors.
      \sys instance is deployed in Europe.
  }
  \label{fig:mirrors}
\end{figure}

\autoref{fig:mirrors} shows that the latency of downloading the metadata index depends on the number and location of mirrors.
\sys returns the metadata index in less than 400 ms for up to five mirrors on the same continent.
In the case of 10 mirrors, \sys returns the metadata index in less than 1.2 seconds.
We observed higher latency when using mirrors located on different continents, mainly due to higher network latency. 
% An average network latency to mirror located in North America or Asia took XX ms and XX ms, respectively.

The last scenario (\emph{All}) shows that the latencies measured when mirrors are evenly distributed across three continents are similar to the latencies measured when using mirrors located only in North America.
It is a result of \sys implementation; \sys contacts the fastest \emph{f + 1} mirrors, and, in case they present different metadata index, it contacts additional mirrors until reaching the quorum (\emph{f + 1} responses are the same).
Therefore, mirrors in Europe and North America were preferred, and \sys latency depends on the slowest selected mirror.

It is the responsibility of the \sys clients to decide on the tradeoff between security and performance.
The experiment shows that even when specifying nine mirrors distributed across different continents, \sys returns the metadata index in about 2.2 seconds. 

    % - hostname: http://uk.alpinelinux.org/alpine/
    %   certificate_chain: |-
    %     UK
    % - hostname: http://dl-5.alpinelinux.org/alpine/
    %   certificate_chain: |-
    %     DE
    % - hostname: http://dl-4.alpinelinux.org/alpine/
    %   certificate_chain: |-
    %     NL
    % - hostname: http://pkg.adfinis-sygroup.ch/alpine/
    %   certificate_chain: |-
    %     CH
    % - hostname: http://dl-8.alpinelinux.org/alpine/
    %   certificate_chain: |-
    %     IT
    % - hostname: http://mirror1.hs-esslingen.de/pub/Mirrors/alpine/
    %   certificate_chain: |-
    %     DE2
    % - hostname: http://mirror.operationtulip.com/alpine/
    %   certificate_chain: |-
    %     DE2
    % - hostname: http://alpine.mirror.didstopia.com/
    %   certificate_chain: |-
    %     DE2
    % - hostname: http://mirror.neostrada.nl/alpine/
    %   certificate_chain: |-
    %     DE2
    % - hostname: http://mirrors.ircam.fr/pub/alpine/
    %   certificate_chain: |-
    %     DE2

%!TEX root = paper.tex
\section{Related work}
\label{sec:related_work}
% in many software-based systems, regarding fixing bugs/vulnerabilities and adding new features, 
Given the importance of software updates, a plethora of works has been proposed to ensure the security of software update systems~\cite{p2p_update_repositories, tuf, chainiac, kshot}. 
Typically, they aim to protect the updates using cryptographic signatures and transfer them to targets via secure connections.
The critical aspect of these approaches is how to protect the signing keys because their leakage compromises the update process. 

The Update Framework (TUF) \cite{tuf} addresses the problem by assigning different roles for accessing specific signing keys, raising the bar for an adversary to get in possession of all keys.
Unfortunately, TUF requires an online project registration; thus it cannot protect a community repository against several attacks, such as delivering arbitrarily modified packages.
Diplomat \cite{kuppusamy_diplomat_2016} overcomes the shortcoming of TUF by dividing signing keys into offline and online keys. The online keys are used to provide fast package signing, a feature required in community repositories. 
Only online keys are leaked in the case of a repository compromise, which is a manageable problem since they can be easily revoked and the repository with new online keys can be regenerated using well-protected offline keys. 
CHAINIAC \cite{chainiac} provides mechanisms to secure the entire software supply chain. Developers create Merkel trees defining software packages with their corresponding binaries. To approve the package release, they sign and submit the trees to co-signing witness servers, which verify the signatures from developers as well as the mapping between the sources and the binaries. This mechanism relies on the blockchain technology, which permits the maintenance of the history of the releases but it increases the system's complexity. 
% Consequently, it is difficult to use CHAINIAC in practice. 
With a similar goal but reduced complexity, {\em in-toto}~\cite{in-toto} offers a mechanism to ensure the integrity of
the software supply chain cryptographically. It enables users with the integrity verification of the whole software supply chain. However, CHAINIAC, {\em in-toto}, and TUF do not consider the case that the target systems are under the protection of trusted computing mechanisms. Thus, they do not protect against integrity violations caused by software updates. 
Recently, KShot~\cite{kshot}  introduced a secure kernel live patching mechanism to fix security vulnerabilities. KShot makes use of system management mode and SGX to perform the patching process without trusting the underlying OS securely. Similarly, \sys leverages SGX to protect the software update patching mechanism (sanitization), but \sys also ensures that software updates do not break the OS integrity. 
We selected Intel SGX to implement \sys since it has become available in clouds~\cite{IBMCloudSGX, AzureSGX}, ported many of confidential cloud native applications including analytics systems~\cite{sgx-pyspark, securetf}, key management system~\cite{palaemon_2020}, and performance monitoring~\cite{teemon}.
% \sys ensures that these updates are safe to be installed in the protected OS.

%  called {\em co-thorities}. These witness servers 

%\wojciech{creating a dedicated mirror with modified packages:}
%Berger et al. 2016 \cite{imasig_updates} partially addressed the problem of providing OS updates that do not break system integrity.
%In \sys, we rely on the proposed idea of building a dedicated mirror serving packages that contain digital signatures.
%Our contribution consist of overcoming the shortcomings with securing the signing key and supporting the installation scripts by leveraging TEE and introducing the concept of sanitization.
% First, \sys proposes how to sanitize packages to allow system reconfigurations caused by custom installation scripts included in packages.
% Second, \sys leverages \gls{tee} to securely generate and protect the signing key.
\sys follows the idea introduced by Berger et al.~\cite{imasig_updates} to maintain custom mirror with modified packages containing digital signatures. 
Unlike the previous work, \sys removes the mirror owner from trusted computing base by protecting the signing keys using TEE.
Also, \sys introduces the sanitization mechanism to enable the installation of packages containing installation scripts.

Several previous studies also considered various security aspects of the mirrors in software update systems~\cite{cappos_look_2008, knockel_mitm_repositories, cappos_stork}. Knockel et al.~\cite{knockel_mitm_repositories} indicated that man-in-the-middle attacks on third-party software are possible for open infrastructures. Fortunately, this can be handled by securing connections using modern TLS instead of outdated SSL technology. The Stork package manager~\cite{cappos_stork} provided mechanisms to handle various attacks from malicious mirrors by dedicating the selective trust to users, \ie, users specify which packages they trust to install. Mercury~\cite{kuppusamy_mercury_2017} addresses the rollback attacks on software packages \cite{cappos_look_2008, bellissimo_secure_software_update_2006} by maintaining a separated signed metafile at the package manager. However, Mercury did not address the problem of the first update in which a package manager cannot ensure the metadata index freshness. \sys tackles this problem by relying on the repository metadata index obtained from the majority of mirrors under the assumption that most mirrors are trustworthy.

\section{Conclusion}
\label{sec:conclusion}
\glsresetall

In this paper, we presented TSR, a trusted software repository, to support secure software updates for integrity-enforced operating systems relying on trusted computing.
TSR is transparent to the existing implementations of package managers and software repositories.
Importantly, it does not require changes to well-established distribution-specific procedures of creating software packages. 

Our implementation supports 99.76\% of the packages available in Linux Alpine main and community repositories. 
It can be hosted on-premises, \eg, in the cloud, while maintaining strong security properties by running inside a \gls{tee}, enabling clients to define custom security policies, and permitting a minority of software repository mirrors to exhibit Byzantine behavior.

%\myparagraph{Software availability} We will make the source code of \sys publicly available.

\smallskip\noindent
\textbf{Acknowledgment.}
We thank our shepherd Professor Hans P. Reiser and the anonymous reviewers for their insightful comments and suggestions as well as Bohdan Trach, Oleksii Oleksenko, Maksym Planeta, Robert Krahn, and Mimi Zohar for their feedback and help.
The research leading to these results has received funding from the Cloud-KRITIS Project and LEGaTO Project (\url{legato-project.eu}), grant agreement No~780681. 
%\clearpage
%\newpage

%\clearpage 
{
	\small
	\bibliographystyle{abbrv}
	\interlinepenalty=10000
	\bibliography{bibliography}
}

\end{document}